\documentclass{jfm}

\usepackage{graphicx}
\usepackage{epstopdf,epsfig}
\usepackage{newtxtext}
\usepackage{newtxmath}
\usepackage{makecell}
\usepackage{multirow}
\usepackage{natbib}
\usepackage{hyperref}

\newcommand\We{\mbox{\textit{We}}}
\newcommand\Oh{\mbox{\textit{Oh}}}

\newcommand{\RomanNumeralCaps}[1]

\newcommand\ff[1]{(\emph{#1})}

\title{Breakup of particle-laden droplets in airflow}

\author{Zhikun Xu\aff{1},
  Tianyou Wang\aff{1,2}
 \and Zhizhao Che\aff{1,2}
\corresp{\email{chezhizhao@tju.edu.cn}}}
\affiliation{\aff{1}State Key Laboratory of Engines, Tianjin University, Tianjin 300350, China
\aff{2}National Industry-Education Platform of Energy Storage, Tianjin University, Tianjin 300350, China
}

\begin{document}
\maketitle

\begin{abstract}
The atomisation of suspension containing liquid and dispersed particles is prevalent in many applications. Previous studies of droplet breakup mainly focused on homogeneous fluids, and the heterogeneous eﬀect of particles on the breakup progress is unclear. In this study, the breakup of particle-laden droplets in airflow is investigated experimentally. Combining synchronised high-speed images from the side view and the 45$^\circ$ view, we compare the morphology of particle-laden droplets with that of homogeneous fluids in different breakup modes. The results show that the higher effective viscosity of particle-laden droplets affects the initial deformation, and the heterogeneous effect of particles appears in the later breakup stage. To evaluate the heterogeneous effect of particles quantitatively, we eliminate the effect of the higher effective viscosity of particle-laden droplets by comparing cases corresponding to the same inviscid Weber number. The quantitative comparison reveals that the heterogeneous effect of particles accelerates the fragmentation of liquid film and promotes localised rapid piercing. A correlation length that depends on the particle diameter and the volume fraction is proposed to characterise the length scale of the concentration fluctuation under the combined effect of the initial flattening and later stretching during the droplet breakup process. Based on this correlation length, the fragment size distributions are analysed, and the scaling results agree well with the experimental data.
\end{abstract}



\section{Introduction}\label{sec:1}

The atomisation of suspension is a fundamental issue for many natural phenomena and practical applications, such as slurry-fuel atomisation \citep{Gvozdyakov2021Slurries, Kim2022Slurry}, spray coating \citep{Koivuluoto2022Coating, Upadhyay2020FoodFilm, Huang2022Coating}, respiratory drug delivery \citep{Andreas2016Pharmatechnology} and ocean spray \citep{Veron2015OceanSpray}. In these processes, the fluids contain dispersed objects such as metal or coal particles, polymer or catalyst, and pollutant plastics, which complicate their atomisation processes. Fluid atomisation usually involves several stages, of which droplet breakup in the airflow is an important process. The droplet breakup produces numerous small fragments, and the fragment size is affected by the dispersed particles, which is crucial to the subsequent heat/mass transfer and reaction. For example, the breakup of slurry fuel droplets directly affects the ignition and combustion of the fuel \citep{Manisha2021Gel}. In addition, the interaction between the fluid and the airflow during the droplet breakup process also exists in primary breakup processes. Therefore, the study of the breakup of particle-laden droplets is useful for understanding the atomisation process of suspensions.

Droplet breakup of homogeneous fluids has been studied by many researchers \citep{Guildenbecher2009SecondaryAtomization, Sharma2022BreakupReview, Theofanous2011DropBreakup}. The breakup process is mainly controlled by the aerodynamic force of the airflow, and the viscosity and surface tension of the liquid. The aerodynamic force of the airflow drives the deformation and breakup of the droplet. As the aerodynamic force increases, droplets in the airflow exhibit different breakup morphologies \citep{Sharma2022BreakupReview, Theofanous2011DropBreakup}, including bag breakup, multimode breakup, shear-stripping breakup, etc. In contrast, the viscosity of the droplet inhibits the droplet deformation, resulting in a stronger aerodynamic force required for high-viscosity fluids to achieve the same breakup morphology \citep{Radhakrishna2021HighOh, Xu2023Viscosity}. The magnitude of the aerodynamic force of the airflow and the viscous force of the droplet relative to the droplet surface tension can be characterised by the Weber number ($\We_g$) and the Ohnesorge number ($\Oh$), respectively. In addition to the breakup morphologies, the droplet breakup process can be quantified with their time characteristics (such as the initial deformation time and the total breakup time \citep{Guildenbecher2009SecondaryAtomization}), kinematic and dynamic characteristics (such as the velocity, acceleration and drag coefficient of the droplet \citep{Theofanous2011DropBreakup, Yang2016HighDensity}), and shape characteristics (such as the droplet width and thickness, the fragment size \citep{Jackiw2021InternalFlow, Jackiw2022SizeDistribution, Anand2022DropletDeformation}).

Due to the multiphase interaction and the complex morphology exhibited during the droplet breakup, the underlying mechanisms of droplet breakup are often interpreted from different perspectives in different studies. From the perspective of surface instability, different breakup morphologies emerge under the competition of the middle piercing controlled by the Rayleigh-Taylor (RT) instability and the peripheral stripping controlled by the Kelvin-Helmholtz (KH) instability \citep{Guildenbecher2009SecondaryAtomization, Sharma2021Aerobreakup, Sharma2022BreakupReview, Theofanous2011DropBreakup}. For the RT instability, in addition to the simplified RT instability being used, the RT instability considering the finite thickness of the liquid was developed for the analysis of the droplet piercing \citep{Theofanous2012ViscousLiquids} and the breakup of the bag film \citep{Villermaux2020FragmentationCohesion, Villermaux2009Raindrops}. For the KH instability, different velocity profiles at the liquid-air interface were discussed and used in the shear-stripping breakup process \citep{Sharma2021Aerobreakup, Sharma2022BreakupReview, Theofanous2012ViscousLiquids}. Moreover, from the perspective of droplet internal flow, the two-point approach \citep{Jackiw2021InternalFlow, Kulkarni2014BagBreakup, Amsden1987TAB} (mathematically consistent with the Taylor analogy breakup (TAB) model) and the energy approach \citep{Ibrahim1993TAB} (i.e., the droplet deformation and breakup (DDB) model) are proposed to describe the droplet deformation and breakup. Different variations and improvements have been developed based on this perspective, but mostly for low Weber numbers \citep{Sojka2021MultimodeBreakup, Rimbert2020DropletDeformation, Stefanitsis2019TAB, Wang2015DDB}. In addition, from the perspective of droplet external airflow, the formation and shedding of vortices outside the droplet are found to significantly affect the droplet breakup at high Weber numbers \citep{Dorschner2020TransverseRTinstability, Liu2018BreakupinSupersonic, Wang2020MachNumber}. Finally, besides these perspectives, other models of droplet deformation and breakup have been proposed based on global principles, including the virtual work principle \citep{Sichani2015DDB} and the maximum entropy formalism \citep{Trujillo2022MEF}.

Different from homogeneous fluids, suspensions exhibit some unique properties that depend primarily on the length scales of the process and the volume fraction of the particles \citep{Foss2007ExperimentalFluidMechanics}. At a length scale much larger than the particle size, the suspension can be treated as a homogeneous fluid. In addition, its rheological properties vary with the volume fraction of the particles, exhibiting Newtonian behaviours at low volume fractions, the shear-thickening or shear-thinning behaviours at high volume fractions, and jamming near the maximum packing fraction \citep{Guazzelli2018Suspensions, More2021Suspensions, Morris2020ConcentratedSuspensions, Ness2022DenseSuspensions}. In contrast, at the length scale close to the particle size, the heterogeneous effects of the particles in the suspension need to be taken into account. \cite{Furbank2004DropletFormation, Furbank2007PinchOff} first reported the pinch-off dynamics of particle-laden liquids and proposed a two-stage description of the droplet formation from suspensions, including the early stage where the suspensions behave as an effective Newtonian fluid and the later stage where the detachment are accelerated caused by the heterogeneous effects of particles. The detailed process, stage transition, and underlying mechanisms of the pinch-off dynamics of particle-laden liquids were further developed by \citep{Bonnoit2012DropDetachment, Lhuissier2018PinchoffSuspension, Clasen2018PinchoffSuspensions, Sauret2022PinchoffSuspensionDrops, Sauret2021PinchoffBidisperseSuspensions}. Similarly, for the breakup of suspension jets, the heterogeneity of the particles affects the shape and the breakup length of the jets \citep{Lhuissier2019SuspensionJet, Stone2011Suspensions}. Moreover, for the spreading and fragmentation of the suspension films, the presence of particles changes the capillary flows of films and leads to more complex situations. \cite{Sauret2020ParticleSheets} experimentally studied the fragmentation of particle-laden liquid sheets generated by a droplet impacting a cylindrical target, and found the presence of particles modifies the thickness and reduces the stability of the liquid sheet. \cite{Jeong2022Dipcoating}, \cite{Gans2019DipCoating}, and \cite{Palma2019DipCoating} reported the dip-coating of a substrate withdrawn from a suspension bath and classified different coating regimes based on the entrainment degree of the particles on the liquid film, which is governed by the film thickness relative to the particle diameter. As for the breakup of the particle-laden droplet considered in this study, the droplet deformation and breakup process under the action of airflow exhibits complex morphologies, including films, ligaments, nodes, etc. For the breakup of these complex morphologies, the variation in length scale will occur as the droplet deforms and breaks up and the effect of particles will be complex. \cite{Zhao2011Slurry} and \cite{Zhao2021ShearThickening} studied the effect of the yield stress and the discontinuous shear thickening exhibited by suspensions on droplet breakup and found some new breakup modes, such as hole breakup, tensile breakup, and hardened deformation. These studies treated suspensions as homogeneous fluids, but the heterogeneous effect of particles on droplet breakup has not been studied.

In this study, the breakup of particle-laden droplets in airflow is investigated. We first discuss the changes in the deformation morphology and breakup mode of particle-laden droplets compared with homogeneous fluids. To quantitatively analyse the effect of particles on the breakup process, we distinguish the effects of the particles via the effective viscosity and via the heterogeneity by comparing at the same breakup mode, i.e., the same inviscid Weber number. Based on this, we further investigate the heterogeneous effects of particles on the typical morphological features and fragment sizes in different breakup modes.

\section{Experimental set-up and materials}\label{sec:2}
\begin{figure}
  \centerline{\includegraphics[width=0.7\columnwidth]{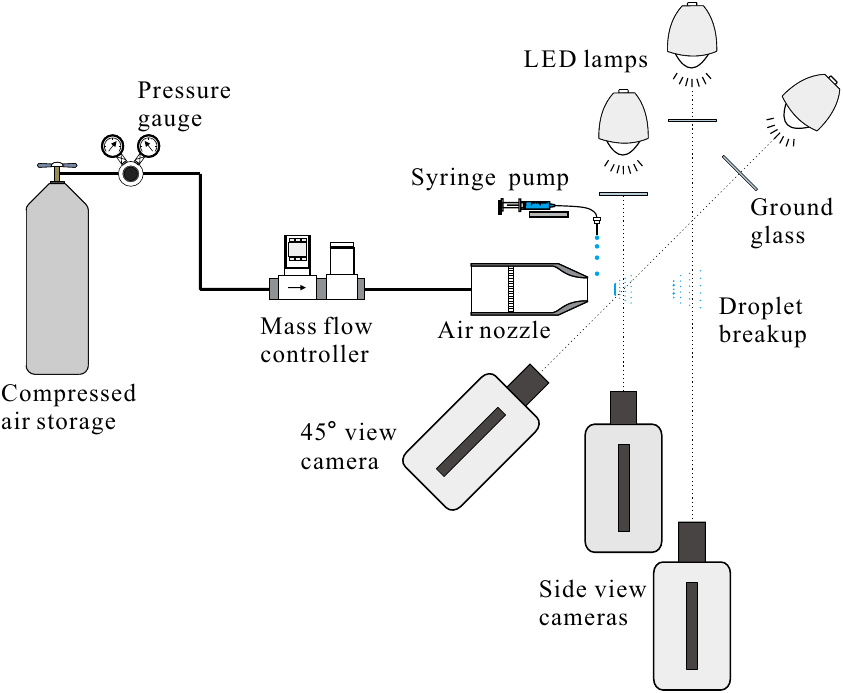}}
  \caption{Schematic diagram of the experimental set-up.}
\label{fig:fig01}
\end{figure}

The experimental set-up is shown in figure \ref{fig:fig01}. A continuous airflow was used to study the droplet breakup, which has been adopted in many studies \citep{Jackiw2021InternalFlow, Radhakrishna2021HighOh, Zhao2021ShearThickening}. The airflow originated from compressed-air storage and was regulated by a mass flow controller (Alicat MCRQ, maximum flow rate 3000 standard litres per minute, estimated uncertainty $\pm$0.8 \% of reading and $\pm$0.2 \% of full scale). The airflow then passed through a honeycomb (a hexagon with a side length of 1 mm) and two-layer mashes (100 mesh) to reduce turbulence, and was finally ejected from a rectangular nozzle (the outlet cross-section is 20 mm in width and 30 mm in height). The range of the airflow velocity was $u_g =$ 8.9--70 m~s$^{-1}$. Droplets entered the airflow from above by gravity. To minimise the eﬀect of the jet shear layer, the falling velocity of the droplet was adjusted to ensure that deformation and breakup of the droplet occurred in the uniform velocity region of the jet for all airflow velocities considered in this study. The experiments were performed at room temperature (22 $^\circ$C) and atmospheric pressure with air density $\rho_g = 1.2$ kg~m$^{-3}$.

High-speed cameras with different arrangements were used to film the droplet deformation and breakup for different purposes. To capture the complete morphology of the droplet breakup, two synchronised cameras (Photron Fastcam SA1.1) were used to take images from the side view and 45$^\circ$ view as illustrated in figure \ref{fig:fig01}. From the side view, a camera with a spatial resolution of 90 $\upmu$m pixel$^{-1}$ acquired the images of the side of the droplet. From the 45$^\circ$ view, a camera with a spatial resolution of 66 $\upmu$m~pixel$^{-1}$ obtained the information on the windward side of the droplet. Both cameras were synchronised at a frame rate of 10,000 frames per second (fps). In addition, to obtain the fragment size after the droplet breakup, two synchronised cameras (Phantom v1612 for the upstream, and Photron Fastcam SA1.1 for the downstream) were arranged together on the side. By combining the shooting areas of the two cameras, we were able to capture all fragments with a sufficient spatial resolution of 60 $\upmu$m~pixel$^{-1}$. Both cameras were set at the frame rate of 5000 fps. For all camera arrangements, macro lenses (Nikon AF 100 mm f/2.8D) with a small aperture (F16 or F22) were used to obtain a large depth of field with low optical distortion, and two high-power (280 W) light-emitting diode (LED) lights diﬀused by ground glasses were used as the background light sources to ensure sufficient brightness for the high-speed imaging. The high-speed images were analysed via a customised image-processing program in Matlab, and the details are given in the supplementary material.

\begin{figure}
  \centerline{\includegraphics[width=0.95\columnwidth]{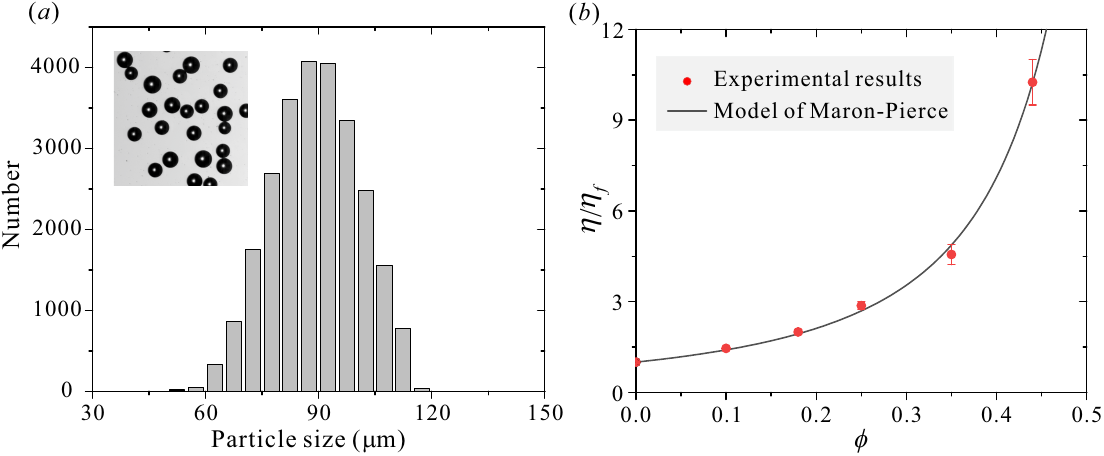}}
  \caption{\ff{a} Particle size distribution of $d_p = 90$ $\upmu$m. The inset is an image of particles under the microscope. \ff{b} Relative effective shear viscosity of the suspensions ($\eta /{{\eta }_{f}}$) for different particle volume fractions ($\phi $). The error bars of the effective viscosity were calculated from the data in the shear rate range of 100--1000 s$^{-1}$.}
\label{fig:fig02}
\end{figure}

The fluids used for droplet breakup were suspensions of solid particles in interstitial liquids. The solid particles were spherical polystyrene particles with a density of $\rho_p = 1050$ kg~m$^{-3}$. The particle sizes were measured from the images taken under an inverted microscope (Nikon Ti-U with a spatial resolution of 2 $\upmu$m~pixel$^{-1}$), and figure \ref{fig:fig02}\ff{a} showed a typical particle size distribution with $d_p = 90$ $\upmu$m. In our experiments, three different particle sizes ($d_p = 40$, 90 and 180 $\upmu$m) were used and their relative standard deviation was about 12\%. The interstitial liquid was silicone oil (phenyl methyl silicone oil from Macklin), which had a viscosity $\eta_f = 35$ mPa~s, a density $\rho_f = 1050$ kg~m$^{-3}$, and a surface tension $\sigma = 27$ mN~m$^{-1}$. The suspensions were repeatedly stirred and rested to remove air bubbles. The particles are totally wetted by the silicone oil and did not modify the liquid-gas interface, so the change of surface tension caused by the particle can be negligible \citep{Lhuissier2018PinchoffSuspension, Jeong2022Dipcoating}. Due to the matching in density between the particles and the interstitial liquid, the effect of particle sedimentation was limited during the experiment.

The effective shear viscosity ($\eta$) of each suspension was measured by a rotating rheometer (Anton Paar MCR 501). The particle volume fraction ($\phi$) of the suspensions was varied from 0 to 44\% in the rheological measurements. The details of the measurement can be found in Appendix \ref{sec:AppA}. Figure \ref{fig:fig02}\ff{b} presents the relative effective shear viscosity ($\eta /{{\eta }_{f}}$) of the suspensions for different particle volume fractions, indicating that the relative effective shear viscosity can be described well by the Maron-Pierce model \citep{Guazzelli2018Suspensions}
\begin{equation}\label{eq:01}
 \frac{\eta }{{{\eta }_{f}}}={{\left( 1-\frac{\phi}{{\phi }_{m}} \right)}^{-2}},
\end{equation}
where ${{\phi }_m}$ is the maximum packing fraction and ${{\phi }_{m}}=0.64$ in our experiment. To separate the influence of the effective viscosity from the heterogeneous effects of the particles, we chose silicone oils of different viscosities (Phenylmethyl silicone oil of $\eta $ = 35, 100, 200 mPa~s from Macklin) as a control group, which had a surface tension $\sigma = 27 \pm 1$ mN~m$^{-1}$. The particle volume fraction ($\phi $) in the droplet breakup experiments was varied from 0 to 35\% because the droplets tend to be non-spherical at high particle volume fractions. The droplet diameter used in this study was $d_0 = 3.35 \pm 0.05$ mm. Correspondingly, the range of the gas Weber number (${{\We}_{g}}={{\rho }_{g}}u_{g}^{2}{{d}_{0}}/\sigma $) in this study was 11.8--718, and the range of the Ohnesorge number ($\Oh=\eta /\sqrt{{{\rho }_{f}}{{d}_{0}}\sigma }$) was 0.11--0.65.

\section{Results and discussion}\label{sec:3}
\subsection{Droplet morphology during the deformation and breakup}\label{sec:3.1}
\subsubsection{Bag and bag-stamen breakup}
\begin{figure}
  \centerline{\includegraphics[width=0.75\columnwidth]{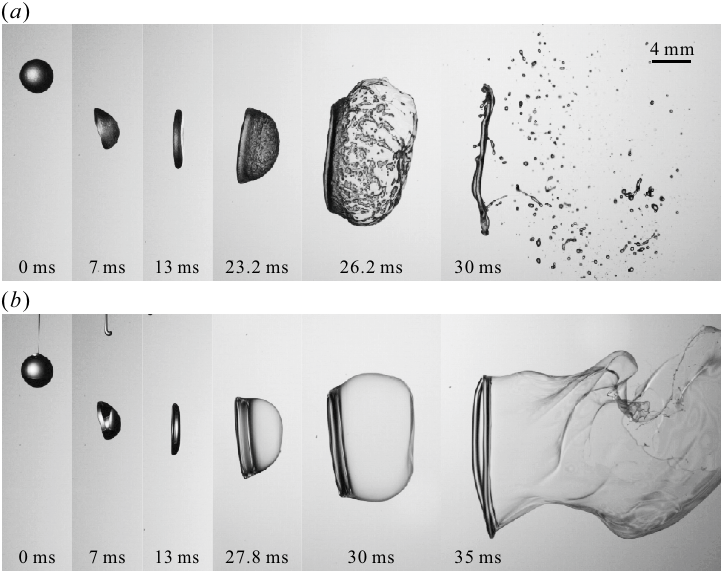}}
  \caption{Image sequences of the bag breakup mode at $\We_g = 20.3$, $\Oh = 0.32$: \ff{a}~suspension with $d_p = 90$ $\upmu$m and $\phi = 0.25$; \ff{b} silicone oil with $\eta = 100$ mPa s. The corresponding video can be found in the supplementary material (Movie 1).}
\label{fig:fig03}
\end{figure}

The bag breakup mode is one of the main modes of droplet breakup. Figure \ref{fig:fig03}\ff{a} shows the bag breakup process of a particle-laden droplet. For comparison, the bag breakup process of a silicone oil droplet is shown in figure \ref{fig:fig03}\ff{b}, which has the same effective viscosity as the particle-laden droplet. The breakup process can be divided into an initial deformation stage (0--13 ms in figure \ref{fig:fig03}), and a bag development and breakup stage (after 13 ms in figure \ref{fig:fig03}). During the initial deformation stage, the particle-laden droplet deforms almost the same as the silicone oil droplet. This indicates that during the initial deformation stage, the particle-laden droplet can be regarded as a homogeneous fluid; and compared to the interstitial liquid, the particles lead to a higher effective viscosity of the droplet. However, during the bag development and breakup stage, the heterogeneous effect of particles appears. The particles agglomerate on the liquid film and accelerate the breakup of the liquid film (23.2--30 ms in figure \ref{fig:fig03}\ff{a}). In contrast, the liquid film of the silicone oil droplet is significantly stretched and deformed, and the breakup starts later (35 ms in figure \ref{fig:fig03}\ff{b}). The development and breakup of the suspension bag film are similar to the fragmentation of particle-laden liquid sheets generated by a droplet impacting a cylindrical target described by \cite{Sauret2020ParticleSheets}. However, compared with the particle-laden liquid sheets described by \cite{Sauret2020ParticleSheets}, the particle distribution on the suspension bag film is more uneven due to the accelerated stretching of the bag film caused by the continuous aerodynamic force. Further quantitative comparisons for the stretch and breakup of the liquid film will be given in \S \ref{sec:3.3}.

\begin{figure}
  \centerline{\includegraphics[width=0.95\columnwidth]{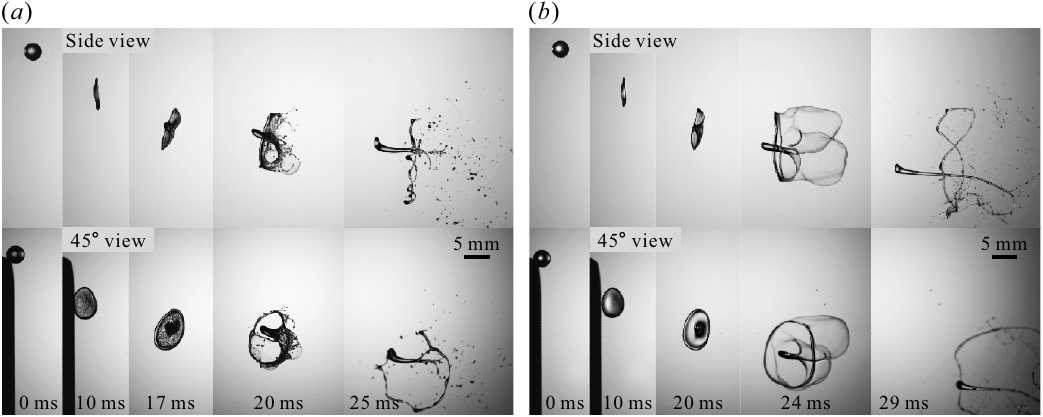}}
  \caption{Synchronised images of the bag-stamen breakup at $\We_g = 28.7$, $\Oh = 0.32$: \ff{a} suspension with $d_p = 90$ $\upmu$m and $\phi = 0.25$, \ff{b} silicone oil with $\eta = 100$ mPa s. The corresponding video can be found in the supplementary material (Movie 2).}
\label{fig:fig04}
\end{figure}

As $\We_g$ increases, the bag-stamen breakup mode occurs. Different from the bag breakup, the droplet morphology in bag-stamen breakup mode consists of the peripheral bag and the middle stamen. The comparison of the breakup process of the particle-laden droplet and the silicone oil droplet is shown in figure \ref{fig:fig04}. The initial deformation of the particle-laden droplet (0--10 ms in figure \ref{fig:fig04}\ff{a}) is similar to that of the silicone oil droplet under the same conditions (figure \ref{fig:fig04}\ff{b}). But after the initial deformation, compared with that of the silicone oil droplet, the peripheral bag of the particle-laden droplet breaks up quickly (20 ms in figure \ref{fig:fig04}\ff{a}), and the elongation of the peripheral ring is weaker (25 ms in figure \ref{fig:fig04}\ff{a}).

\subsubsection{Multimode breakup}\label{sec:3.1.2}
\begin{figure}
  \centerline{\includegraphics[width=1\columnwidth]{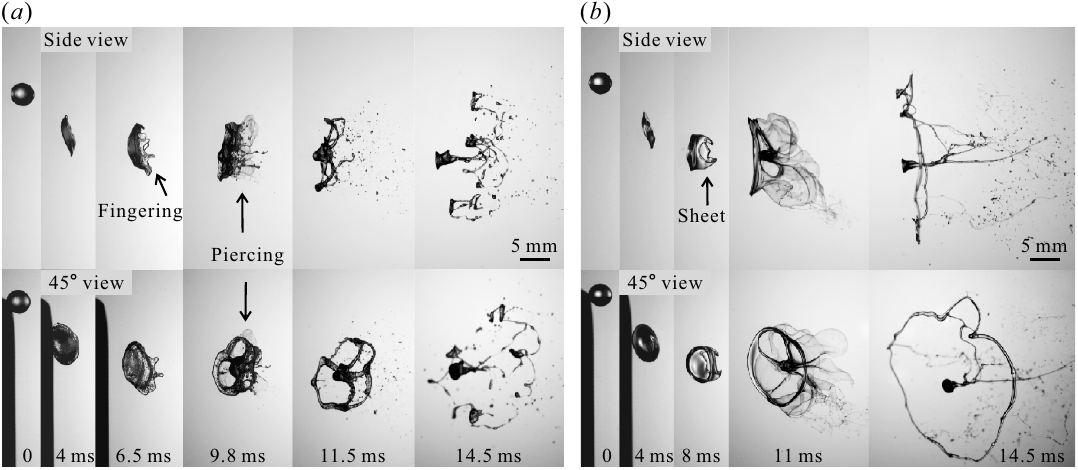}}
  \caption{Synchronised images of the low-order multimode breakup at $\We_g = 71.7$, $\Oh = 0.32$: \ff{a} suspension with $d_p = 90$ $\upmu$m and $\phi = 0.25$; \ff{b} silicone oil with $\eta = 100$ mPa s. The corresponding video can be found in the supplementary material (Movie 3).}
\label{fig:fig05}
\end{figure}

\begin{figure}
  \centerline{\includegraphics[width=0.95\columnwidth]{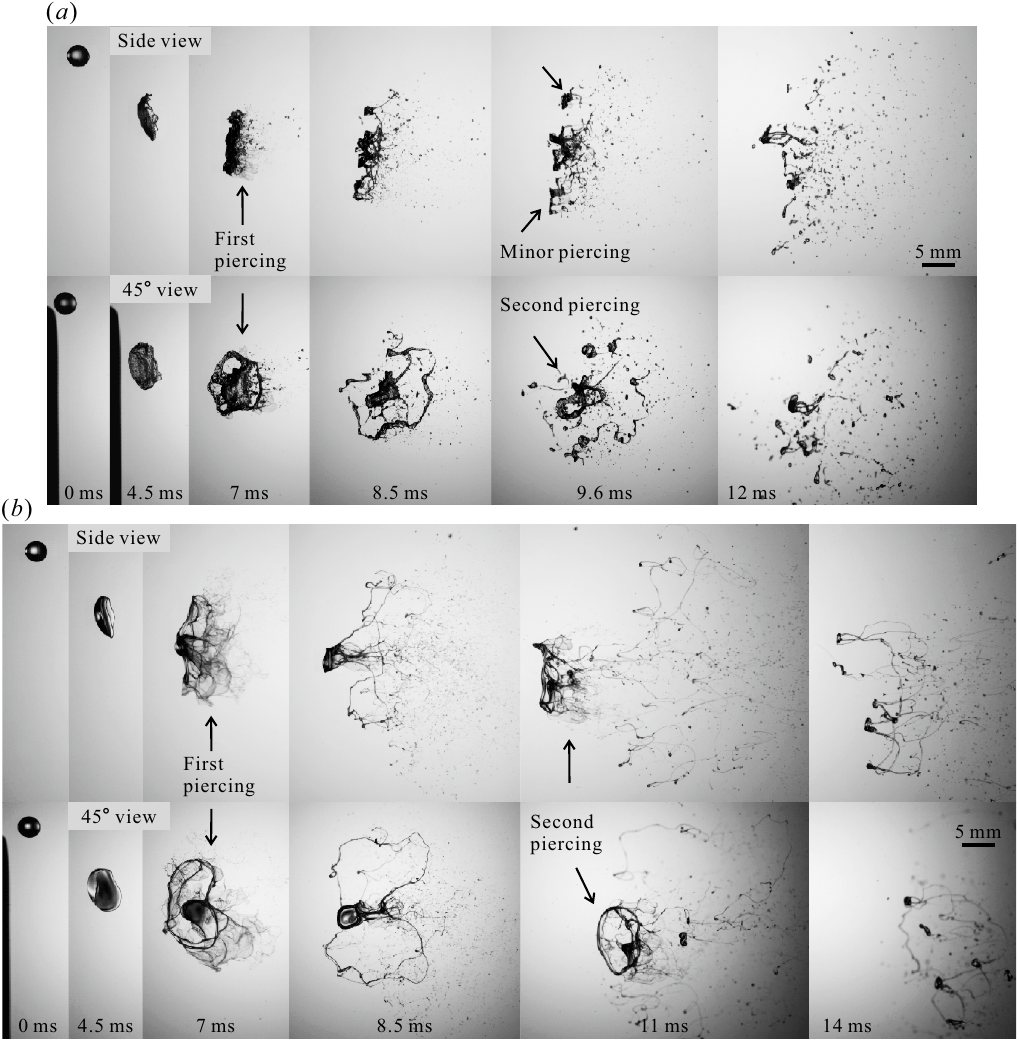}}
  \caption{Synchronised images of the high-order multimode breakup at $\We_g = 103.7$, $\Oh = 0.32$: \ff{a} suspension with $d_p = 90$ $\upmu$m and $\phi = 0.25$, \ff{b} silicone oil with $\eta = 100$ mPa s. The corresponding video can be found in the supplementary material (Movie 4).}
\label{fig:fig06}
\end{figure}

The multimode breakup mode is another main mode of droplet breakup. Compared with bag breakup, the multimode breakup occurs at a higher Weber number and has new features. \cite{Xu2023Viscosity} classified the multimode breakup into low-order and high-order modes according to the piercing times in the middle of the droplet. A typical process of the low-order multimode breakup is shown in figure \ref{fig:fig05}, while the morphology of the high-order multimode breakup is shown in figure \ref{fig:fig06}.

For comparison, figure \ref{fig:fig05}\ff{a} shows the breakup process of a particle-laden droplet and figure \ref{fig:fig05}\ff{b} shows the breakup process of a silicone oil droplet under the same conditions. The process of the low-order multimode breakup can be divided into the initial deformation stage (0--4 ms in figure \ref{fig:fig05}), the peripheral retraction stage (4--6.5 ms in figure \ref{fig:fig05}\ff{a} and 4--8 ms in figure \ref{fig:fig05}\ff{b}), and the subsequent piercing and breakup stage. During the initial deformation stage, the deformation is similar for both fluids, and the particle-laden droplet can be regarded as a homogeneous fluid. During the peripheral retraction stage, the silicone oil droplet tends to form a peripheral sheet, and the peripheral sheet deflects to the leeward of the droplet before the sheet destabilization develops, i.e., the sheet retraction (8 ms in figure \ref{fig:fig05}\ff{b}). In contrast, for the particle-laden droplet under the same conditions, the azimuthal destabilization of the droplet periphery develops and forms fingerings at the periphery (6.5 ms in figure \ref{fig:fig05}\ff{a}). Hence, the formation of fingers indicates the emergence of particle heterogeneity as the peripheral sheet becomes thinner. In addition, for the piercing in the middle of the droplet, the particle-laden droplet tends to form local piercing with a faster speed and less lateral expansion (9.8 ms in figure \ref{fig:fig05}\ff{a}) than the silicone oil droplet (11 ms in figure \ref{fig:fig05}\ff{b}).

\begin{figure}
  \centerline{\includegraphics[width=0.95\columnwidth]{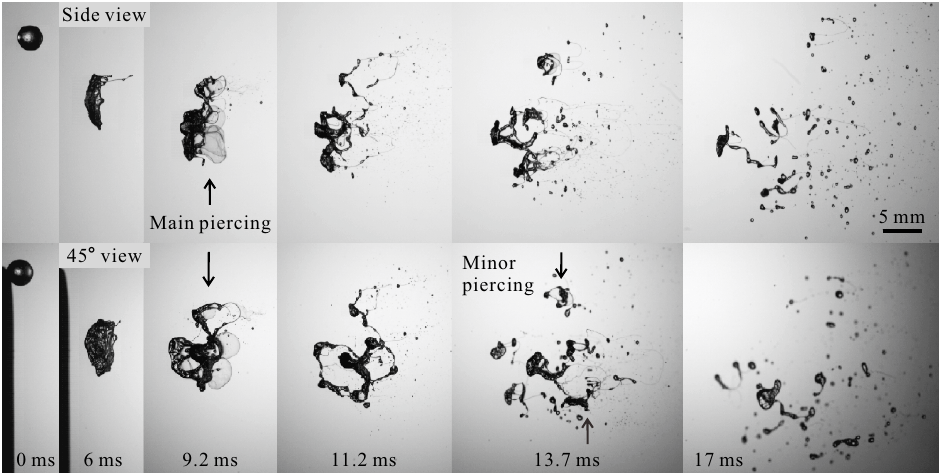}}
  \caption{Synchronised images of the low-order multimode breakup at $\We_g = 69.9$, $\Oh = 0.32$. The fluid used is suspension with $d_p = 180$ $\upmu$m and $\phi = 0.25$. The corresponding video can be found in the supplementary material (Movie 5).}
\label{fig:fig07}
\end{figure}

Different from the low-order multimode breakup where the droplet middle is pierced once, the droplet middle is pierced multiple times under the high-order multimode breakup. The comparison of the particle-laden droplet and the silicone oil droplet is shown in figure \ref{fig:fig06}. The time interval between two piercings of the particle-laden droplet (7--9.6 ms in figure \ref{fig:fig06}\ff{a}) is shorter than that of the silicone oil droplet (7--11 ms in figure \ref{fig:fig06}\ff{b}). In addition to the major piercings, the particle-laden droplet forms some minor piercings (9.6 ms in figure \ref{fig:fig06}\ff{a}). Similar minor piercings also occur in the case of the low-order multimode breakup with large particles (9.2 ms in figure \ref{fig:fig07}). These piercings in the middle have important effects on the breakup morphology and fragment size, which will be discussed further in \S \ref{sec:3.4}.

\subsubsection{Shear-stripping breakup}\label{sec:313}

\begin{figure}
  \centerline{\includegraphics[width=0.8\columnwidth]{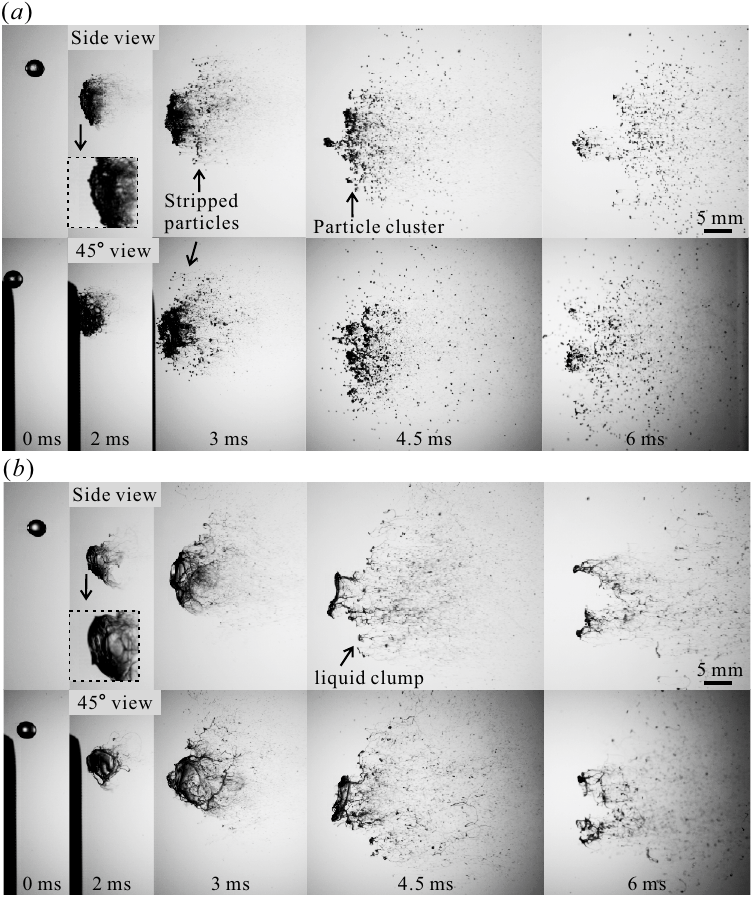}}
  \caption{Synchronised images of the shear-stripping breakup at $\We_g = 459.5$, $\Oh = 0.32$: \ff{a} suspension with $d_p = 180$ $\upmu$m and $\phi = 0.25$; \ff{b} silicone oil with $\eta = 100$ mPa s. The insets show enlarged images of the windward surface of the droplets. The corresponding video can be found in the supplementary material (Movie 6).}
\label{fig:fig08}
\end{figure}

The shear-stripping breakup occurs at a much higher Weber number and is the terminal regime considered by some studies \citep{Theofanous2011DropBreakup, Theofanous2012ViscousLiquids}. Figure \ref{fig:fig08}\ff{a} shows the shear-stripping breakup process of a particle-laden droplet, and figure \ref{fig:fig08}\ff{b} shows the breakup process of a silicone oil droplet under the same conditions for comparison. Being different from the bag and multimode breakup modes, here the heterogeneous effect of particles appears at the beginning of the breakup process. In the early stage of the breakup of the particle-laden droplet (0--3 ms in figure \ref{fig:fig08}\ff{a}), the interstitial liquid is rapidly pushed downstream by the airflow, thus particles protrude from the windward side in the middle of the droplet forming a rough surface, while some particles are stripped directly from the periphery of the droplet. In contrast, for the silicone oil droplet, the windward side in the middle is relatively smoother due to the tension generated by the outward stripping (2 ms in figure \ref{fig:fig08}\ff{b}), while some ripples on the edge of the windward surface of the silicone oil droplet are formed. After the flattening of the droplet middle, massive piercings occur in the middle of the droplet. The particle-laden droplet forms some large particle clusters (4.5 ms in figure \ref{fig:fig08}\ff{a}), while the silicone oil droplet forms some liquid clumps (4.5 ms in figure \ref{fig:fig08}\ff{b}). Finally, the particle clusters or the liquid clumps may be further pierced by the airflow. The particles and the interstitial liquid in the particle-laden droplet are almost completely separated (6 ms in figure \ref{fig:fig08}\ff{a}), while the silicone oil droplet forms mists of tiny droplets (6 ms in figure \ref{fig:fig08}\ff{b}). Overall, compared with the continuous breakup of the silicone oil droplet, the breakup of the particle-laden droplet is more discrete, like the shattering of a solid.

\subsection{Regime map of droplet breakup}\label{sec:3.2}

The morphology of droplet breakup is a complex process with different breakup patterns in different conditions. Here, according to the criteria for viscous droplets \citep{Xu2023Viscosity} (i.e., based on the piercing time in the droplet middle and the retraction in the droplet periphery), we classify the droplet breakup modes into no-breakup, bag breakup (figure \ref{fig:fig03}), bag-stamen breakup (figure \ref{fig:fig04}), low-order multimode breakup (figure \ref{fig:fig05}), high-order multimode breakup (figure \ref{fig:fig06}), and shear-stripping breakup (figure \ref{fig:fig08}). The transition from bag breakup to bag-stamen breakup corresponds to the formation of a stamen at the middle of the droplet (24 ms in figure \ref{fig:fig04}\ff{b}). The transition from bag-stamen breakup to low-order multimode breakup corresponds to the formation of a sheet at the periphery of the droplet (8 ms in figure \ref{fig:fig05}\ff{b}) and the piercing at the middle of the droplet (11 ms in figure \ref{fig:fig05}\ff{b}). The transition from low-order multimode breakup to high-order multimode breakup corresponds to the change of the piercing times in the droplet middle. After the sheet formation at the periphery, the middle part of the droplet is pierced once in the low-order multimode breakup mode (figure \ref{fig:fig05}) and multiple times in the high-order multimode breakup mode (figure \ref{fig:fig06}). Finally, the transition from high-order multimode breakup to shear-stripping breakup corresponds to the occurrence of peripheral stripping (figure \ref{fig:fig08}), i.e., the appearance of Kelvin-Helmholtz instability waves at the periphery of the droplet. The main morphology features of different breakup modes are summarised in table \ref{tab:tab1}.

\begin{table}
\begin{center}
\def~{\hphantom{0}}
\begingroup
\setlength{\tabcolsep}{18pt}
\begin{tabular}{lll}
\textbf{Breakup mode}         & \textbf{Middle of the droplet} & \textbf{Periphery of the droplet} \\
No-breakup           & \multicolumn{2}{l}{Deformation and oscillation}  \\
Bag breakup          & \multicolumn{2}{l}{Bag with a toroidal rim}      \\
Bag-stamen           & Stamen                & Peripheral bag           \\
Low-order multimode  & Single piercing       & Peripheral bag or sheet  \\
High-order multimode & Multiple piercings    & Peripheral sheet         \\
Shear-stripping      & Massive piercings     & Stripping\\
\end{tabular}%
\endgroup
\caption{Main morphological features of different breakup modes.}
\label{tab:tab1}
\end{center}
\end{table}

\begin{figure}
  \centerline{\includegraphics[width=0.75\columnwidth]{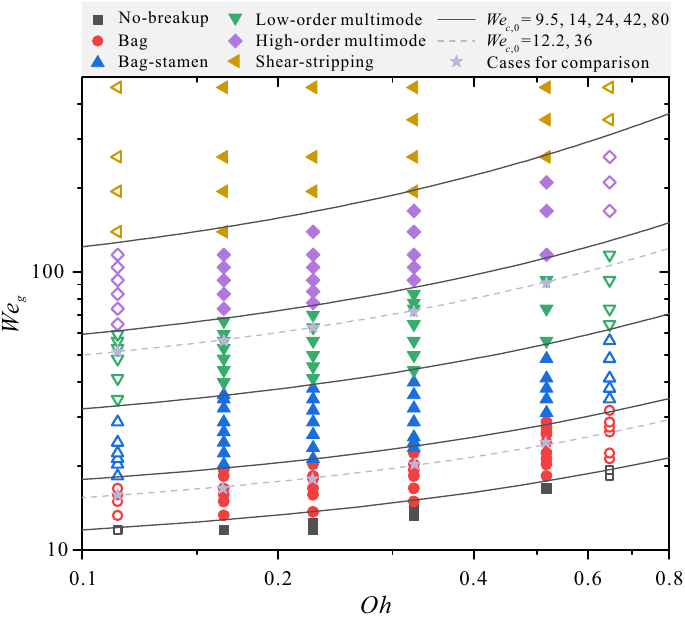}}
  \caption{Regime map of droplet breakup. The hollow symbols correspond to the breakup of silicone oil droplets and the solid symbols correspond to the breakup of particle-laden droplets. For particle-laden droplets, $\Oh$ is calculated based on the effective viscosity (\ref{eq:01}) at the corresponding particle volume fraction. The solid lines are based on the theoretical model (\ref{eq:02}) for the transitions of diﬀerent regimes. The stars along the dashed lines correspond to the cases for the comparison under the same $\We_{c,0}$.}
\label{fig:fig09}
\end{figure}

The regime map is shown in figure \ref{fig:fig09}. The transition of different modes is determined by the comparison between the droplet width after the initial flattening ($d_w$) and the most-amplified wavelength of the RT instability (${\lambda}_{RT}$). ${\lambda}_{RT}$ was obtained by \cite{Aliseda2008ViscousAtomization}, and $d_w$ can be obtained by balancing the kinetic energy of the airflow and the surface energy of the droplet. By linking these processes, a theoretical model was proposed by \cite{Zhao2011BagBreakup} and further developed by \cite{Xu2023Viscosity} for the transitions between different breakup modes
\begin{equation}\label{eq:02}
 {{\left( \frac{{{\We}_{c,0}}}{{{\We}_{c}}} \right)}^{1/2}}+C\We_{c,0}^{2/3}{{\left( \frac{{{\Oh}^{2}}}{{{\We}_{c}}} \right)}^{1/3}}=1,
\end{equation}
where $\We_c$ is the critical Weber number for the transition between different regimes. $C$ is equal to 0.24 and is universal for $Oh<2.15$, which can be considered as a weight coefficient of the viscous effect relative to the surface tension in the droplet breakup process \citep{Xu2023Viscosity}. When $\Oh$ approaches zero, $\We_c$ is equal to the critical Weber number in the inviscid case ($\We_{c,0}$). Under the same $\We_{c,0}$, the degree of initial deformation of the droplet is similar, and the inhibition of the initial deformation by the higher viscosity of the droplets is counteracted by a higher Weber number \citep{Cohen1994ViscousBreakup}. Here, we use this model to obtain the transitions of different breakup modes, as shown by the solid lines in figure \ref{fig:fig09}. The critical Weber number (${We}_c$) of both the suspension droplets and the silicone droplets is obtained through (\ref{eq:02}).

Comparing the breakup modes of silicone oil droplets (the hollow symbols in figure \ref{fig:fig09}) and particle-laden droplets (the solid symbols in figure \ref{fig:fig09}), we can see that the effect of particles on the transition of breakup mode is manifested by a higher viscosity compared to the interstitial fluid. It is because the transition of the breakup mode is controlled by the interfacial instability underlying the initial deformation. For the initial flattening deformation, the suspension can be regarded as a homogeneous fluid, as described in \S \ref{sec:3.1}. For the interfacial instability, the instability wavelength (${{\lambda }_{RT}}\sim {{d}_{0}}$) is much larger than the particle size, so the heterogeneous effect of particles on interfacial instability is minor. Therefore, the breakup mode can be classified based on the effective viscosity of the particle-laden droplet, and the heterogeneous effect of particles appears only in the later breakup stage.

Since the addition of particles into the liquid can increase the effective viscosity of the fluid, the contribution of the higher viscosity should be eliminated when we analyse the heterogeneous effect of the particles. Otherwise, the effects of the higher viscosity and the effects of the particle heterogeneity are mixed, hindering us from understanding the underlying mechanisms. In this regard, suspensions with different concentrations should be compared under the same $\We_{c,0}$ in (\ref{eq:02}). (\ref{eq:02}) is to obtain the transitions between different breakup modes, but the transitions between different breakup modes are essentially the change of breakup morphology. Therefore, (\ref{eq:02}) with different $\We_{c,0}$ can reflect the change of breakup morphology, that is, the droplet has a similar breakup morphology under the condition obtained by (\ref{eq:02}) with a certain $\We_{c,0}$. In this way, we can compare the conditions with different particle volume fractions and isolate the effects of particle heterogeneity. In the subsequent sections, we choose two typical breakup morphologies for comparison, i.e., the bag breakup (${{\We}_{c,0}}=12.2$) and the low-order multimode breakup (${{\We}_{c,0}}=36$), as shown by the dashed lines in figure \ref{fig:fig09}.

\subsection{Stretching and fragmentation of liquid film}\label{sec:3.3}
\begin{figure}
  \centerline{\includegraphics[width=0.7\columnwidth]{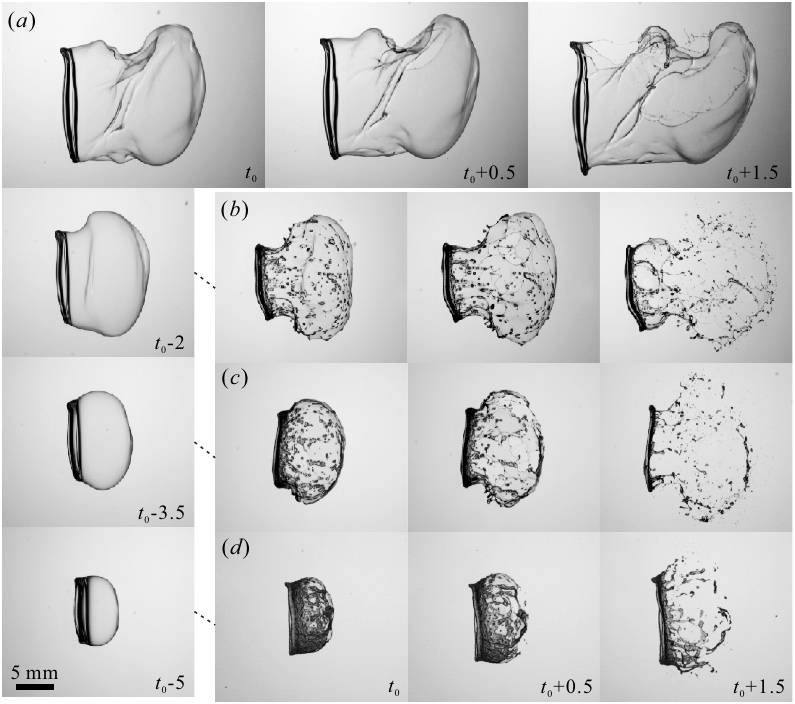}}
  \caption{Image sequences of bag fragmentation processes. The cases correspond to the stars along the dashed line with $\We_{c,0} = 12.2$ in figure \ref{fig:fig09}: \ff{a} silicone oil with $\eta = 35$ mPa~s, $\We_g = 15.7$, $\Oh = 0.11$; \ff{b} suspension with $\phi = 0.10$, $\We_g = 16.6$, $\Oh = 0.17$; \ff{c} suspension with $\phi = 0.25$, $\We_g = 20.3$, $\Oh = 0.32$; \ff{d} suspension with $\phi = 0.35$, $\We_g = 24.3$, $\Oh = 0.52$. The size of the particles within the suspensions is $d_p = 90$ $\upmu$m. $t_0$ corresponds to the moment when the bag fragmentation begins. The images with similar morphology are connected by dashed lines. For comparison with other panels of the figures, the time sequence in \ff{a} is from the bottom left to the top left and then to the top right.}
\label{fig:fig10}
\end{figure}

The stretching and fragmentation of the liquid film is the main morphological feature of droplet breakup, especially for the bag and bag-stamen breakup modes. Here, we compare the fragmentation process of the bag film under the same $ {{\We}_{c,0}}=12.2$, as shown in figure \ref{fig:fig10}. It can be seen that as the particle volume fraction increases, the bag fragmentation occurs earlier in the stretching process. For example, the morphology corresponding to the fragmentation moment ($t_0$) in figure \ref{fig:fig10}\ff{c} is similar to the morphology corresponding to 3.5 ms before the fragmentation moment ($t_0-3.5$) in figure \ref{fig:fig10}\ff{a}. This is because particles occupy a part of the droplet, so the liquid content in the droplet is reduced compared with the case without particles. In addition, the liquid enclosed in particle clusters is not available for the stretching of the liquid film, which is similar to the observation of \cite{Sauret2020ParticleSheets} in the fragmentation of particle-laden liquid sheets. Therefore, the liquid used for the bag stretching is less, which is not conducive to the stretching of the bag film. Moreover, when the thickness of the bag film is smaller than the diameter of the particle clusters, capillary stress will form around the particle clusters. The local stress induced by the particles makes the liquid film susceptible to fragmentation \citep{Villermaux2020FragmentationCohesion}. In addition, the fragmentation processes of the bag film are different. For the silicone oil droplet, the fragmentation of the liquid film starts by forming holes at one or several locations where the film is the thinnest, and then the holes expand. In this process, the kinematic thinning leads to the film rupture \citep{Villermaux2020FragmentationCohesion}. However, for the particle-laden droplet, many holes are initiated on the liquid film due to the presence of the particles, forming a web-like fragmentation. This fragmentation process of the liquid film is the spontaneous hole formation caused by internal defects, i.e., the local stress induced by the particles \citep{Villermaux2020FragmentationCohesion}.

\begin{figure}
  \centerline{\includegraphics[width=1\columnwidth]{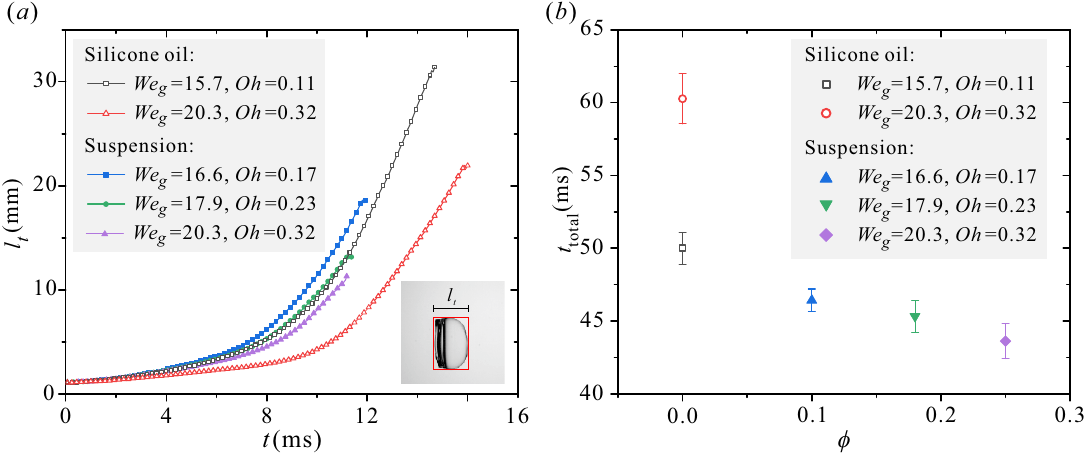}}
  \caption{\ff{a} Evolution of the droplet thickness along the direction of airflow in the bag development stage. The time here is defined from the end of initial deformation (when the droplet is flattened to a disk shape with the largest aspect ratio) to the beginning of bag fragmentation (when holes appear in the bag). \ff{b} Total breakup time under different particle volume fractions. The cases correspond to the stars along the dashed line with $\We_{c,0} = 12.2$ in figure \ref{fig:fig09}, and the particle size in the suspensions is $d_p = 90$ $\upmu$m.}
\label{fig:fig11}
\end{figure}

Through digital image processing, we can obtain the droplet thickness ($l_t$) along the direction of airflow, i.e., the width of the vertical bounding box of the droplet (as shown in the inset of figure \ref{fig:fig11}\ff{a}). The evolution of the droplet thickness in the bag development stage is shown in figure \ref{fig:fig11}\ff{a}. The stretching speed of the liquid film of a particle-laden droplet (solid symbols in figure \ref{fig:fig11}\ff{a}) is closer to that of the interstitial liquid (hollow square symbols in figure \ref{fig:fig11}\ff{a}) than that of liquid with the same effective viscosity (hollow triangular symbols in figure \ref{fig:fig11}\ff{a}). This is because the interaction between the particles is insignificant during the stretching process of the liquid film of the bag, while the viscosity of the interstitial fluid is dominant. Moreover, we calculate the total breakup time, which is the time when all fragmentation has ceased \citep{Pilch1987BreakupTime}, as shown in figure \ref{fig:fig11}\ff{b}. The total breakup time is smaller for particle-laden droplets (solid symbols in figure \ref{fig:fig11}\ff{b}) than for silicone oil droplets (hollow symbols in figure \ref{fig:fig11}\ff{b}) and also decreases slightly with increasing particle volume fraction. This reduction in the total breakup time is mainly because the particles accelerate the fragmentation of the bag (as shown in figure \ref{fig:fig10}\ff{a}) and the ring (similar to the jet breakup described by \cite{Lhuissier2019SuspensionJet}).

\subsection{Piercing by Rayleigh-Taylor instability}\label{sec:3.4}
The piercing by the Rayleigh-Taylor instability wave is another main morphological feature of the droplet. The piercing times in the middle of the droplet that determines the breakup mode are mainly affected by the effective viscosity of the particle-laden droplet, as described in \S \ref{sec:3.2}. However, in the later stages of piercing development, the heterogeneity of the particle-laden droplets becomes important and affects the droplet morphology.

\begin{figure}
  \centerline{\includegraphics[width=0.9\columnwidth]{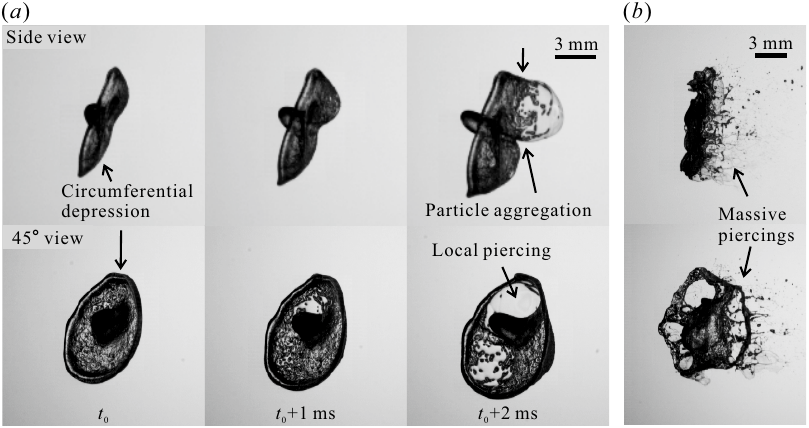}}
  \caption{\ff{a} Evolution of local piercing in the bag-stamen breakup. The whole breakup process is shown in figure \ref{fig:fig04}\ff{a}. \ff{b} First piercing in the high-order multimode breakup. The whole breakup process is shown in figure \ref{fig:fig06}\ff{a}.}
\label{fig:fig12}
\end{figure}

To clearly display the piercing development process, we select the local piercing process of the bag-stamen breakup mode, as shown in figure \ref{fig:fig12}\ff{a}. After the formation of circumferential depression on the windward of the droplet, the local piercing with a faster speed occurs in the interstitial liquid as the droplet becomes thinner. The bag produced by the local piercing contains fewer particles, whereas particles aggregate at the edge of the bag. Inspired by the comparison with the study of \cite{Sauret2020ParticleSheets}, we suggest that the particle aggregation at the junction of the bag film and the ring is caused by the accelerated stretching of the bag film. Let us take the case of figure \ref{fig:fig12}\ff{a} as an example. At the beginning of bag stretching (i.e., the end of droplet flattening), the particle-based Weber number is ${\We_p} = {\rho _p}{d_p}u_f^2/\sigma  \approx 0.77$, where $u_f$ is the velocity of the bag film and can be estimated as ${u_f} \sim {u_g}\sqrt {{\rho _g}/{\rho _f}}  \approx 0.47$ m/s at the beginning of bag stretching. Meanwhile, the Stokes number of the particle is $St = {\rho _p}\dot \gamma d_p^2/{\eta _f} \approx 0.136$, where $\dot \gamma $ is the shear rate of the droplet and can be estimated $\dot \gamma  \sim {u_f}/{d_0}$. So at the early stage of the bag development, the particle motion is in the viscous regime ($St \ll 1$), and the particles can follow the stretching of bag film (as shown at $t_0$ in figure \ref{fig:fig12}\ff{a}). However, as the bag film stretches and becomes thinner, the velocity and the shear rate of the bag film increase rapidly (${u_f} \approx 4.3$ m/s and $\dot \gamma  \sim {u_f}/{d_p}$ at $t_0 + 1$ ms in figure \ref{fig:fig12}\ff{a}) due to the continuous aerodynamic force. ${\We}_p$ and $St$ increase rapidly and can be expected to exceed 1. The motion of the particle shifts to an inertia-dominated regime gradually, and the ability of the particles to follow the stretching of bag film weakens. Therefore, the particles are retained at the junction of the bag film and the ring to form particle aggregation. Moreover, as the particles are retained at the junction, a viscosity difference develops between the tip and the root of the bag, which further increases the velocity difference. This forms a mutually reinforcing process of particle aggregation at the root of the bag film and acceleration at the tip of the bag film. Finally, a local piercing with a faster speed forms (as shown $t_0 + 2$ ms in figure \ref{fig:fig12}\ff{a}). The piercing is faster due to the lower viscosity of the interstitial liquid compared with the effective viscosity of the suspension. Meanwhile, the piercing is localised due to particle aggregation at the bag edge (highlighted by the arrows at $t_0 + 2$ ms in figure \ref{fig:fig12}\ff{a}) inhibiting the lateral expansion of the bag. In addition, another piece of evidence that can reflect the effect of the stretching acceleration of the bag film is that with the increase of the stretching acceleration of the bag film (i.e., increasing $\We_g$, from figure \ref{fig:fig03}\ff{a} to figure \ref{fig:fig05}\ff{a}), the number of particles on the bag film decreases. This indicates that the ability of the particles to follow the stretching of the bag film weakens with the increase of the stretching acceleration of the bag film. As $\We_g$ increases, local piercings become more abundant, as shown in figures \ref{fig:fig05}\ff{a} and \ref{fig:fig12}\ff{b}.

The development of local piercing for particle-laden droplets will produce some new features. For the homogeneous liquid, the RT instability wave with the largest growth rate (i.e., the most-amplified wave) dominates the droplet breakup \citep{Sharma2022BreakupReview, Theofanous2011DropBreakup}, and the lateral expansion of the most-amplified wave can suppress the development of instability waves with other wavelengths. However, for the particle-laden droplet, the most-amplified wave tends to further produce local piercing, while its lateral extension weakens. Therefore, the instability waves with other wavelengths may develop and form some minor piercings, such as 13.7 ms in figure \ref{fig:fig07} and 9.6 ms in figure \ref{fig:fig06}\ff{a}. In addition, in the high-order multimode breakup (figure \ref{fig:fig06}), the time interval between two piercings of the particle-laden droplet (7--9.6 ms in figure \ref{fig:fig06}\ff{a}) is shorter than that of the silicone oil droplet (7--11 ms in figure \ref{fig:fig06}\ff{b}), which also indicates that the most-amplified wave of the particle-laden droplet has a weaker suppression effect on other instability waves. Overall, the piercing phenomenon of the particle-laden droplet is more abundant and random.

\subsection{Analysis of fragment size}\label{sec:3.5}

Understanding and controlling the fragment size of droplet breakup is important for numerous fields \citep{Villermaux2020FragmentationCohesion}. In this section, we will discuss the effect of particles on the fragment size after droplet breakup. The fragment size is measured through post-processing of the images obtained by two synchronised cameras, and the details of the image-processing procedure are given in the supplementary material.

The fragment originates from the rupture of elementary ligaments, which is affected by many factors. For the breakup of silicone oil droplets, the rupture of elementary ligaments is controlled by the Rayleigh-Taylor or Rayleigh-Plateau instabilities, so the instability wavelength determines the breakup length of the ligament and further the fragment size, which has been studied in detail by \cite{Jackiw2022SizeDistribution}. However, for particle-laden droplets, there is a correlation length ($\xi $) that controls the stability of the suspension through a concentration fluctuation. The correlation length can be seen as a characteristic length scale over which a fluctuation will be damped \citep{Sauret2022PinchoffSuspensionDrops}. The scaling relation of the correlation length is different for different flow conditions \citep{Bonnoit2010ScaleDenseSuspensions, Lhuissier2018PinchoffSuspension, Rognon2015DenseGranularFlows, Sauret2022PinchoffSuspensionDrops}. For the breakup of particle-laden droplets, before the final rupture of ligaments, the droplet undergoes initial flattening and then stretching. The correlation length is determined by the combined effects of the flattening and stretching processes. In the initial flattening process, the droplet is sheared strongly. The confined space and rapid shearing increase the particle-particle interaction, producing a condition similar to the crowded condition even at a relatively lower particle volume fraction. In the crowded condition, there exists a typical correlation length ($\xi_1 $) of the velocity field (or the non-affine velocity field in the case of a sheared system), over which particles move cooperatively \citep{Ness2022DenseSuspensions, Teitel2007Jamming}. The range of the correlated motion depends on the collective degrees of freedom in the system, which can be characterised by ${\phi _m} - \phi $. The typical correlation length scale can be found as
\begin{equation}\label{eq:04}
  {{\xi }_{1}}\sim {{d}_{p}}{{\left( {{\phi }_{m}}-\phi  \right)}^{-n}},
\end{equation}
where ${\phi _m}$ is the maximum packing fraction and ${\phi _m} = 0.64$, $n$ is a constant which depends on the properties of the particles and their interactions \citep{Wyart2015ScaleSuspensions}. To determine $n$, we need to know the flow regime of the particles, which depends on the Stokes number ($St$). The range of $St$ in the initial flattening process is about 0.005--0.22, so a viscous regime ($St \ll 1$) is expected \citep{Ness2022DenseSuspensions}. In the viscous regime, \cite{Teitel2007Jamming} reported $n = 0.6$ based on their numerical results. \cite{Wyart2015ScaleSuspensions} predicted $n = 0.43$ based on a scaling analysis and argued that the effect of inter-particle friction is weak in the viscous regime. The exact value of $n$ is obscure, and here, we choose an intermediate value $n = 0.5$.

After the initial flattening, the droplet is stretched to produce fragments. The basic thinning dynamics of the stretching process are similar to that of the pinch-off process of a suspension droplet, that is, the liquid between two portions of suspension is pulled apart. For this basic thinning dynamics process, \cite{Sauret2022PinchoffSuspensionDrops} proposed a dislocation mechanism for the thinning dynamics to reveal the heterogeneity effect of particles. By balancing the power associated with the capillary forces and the viscous dissipation, they obtained a scaling law for the thinning dynamics. The applicable minimum volume fraction of this scaling law ranges from less than 2\% for 20 $\upmu$m particles to 20\% for 500 $\upmu$m particles, which is generally satisfied in our experiments. Here, considering that the cooperative motion of particles in the flattening process further affects the thinning dynamics in the stretching process, the particle diameter in the original model of \cite{Sauret2022PinchoffSuspensionDrops} is replaced with ${{\xi }_{1}}$, that is
\begin{equation}\label{eq:06}
 \frac{\xi }{{{\xi }_{1}}}\sim {{\left( \frac{{{l}_{c}}}{{{\xi }_{1}}} \right)}^{2/3}}{{\left( 1-\frac{\phi }{{{\phi }_{m}}} \right)}^{-2/3}},
\end{equation}
where ${{\xi }_{1}}$ is the size of the particle cluster moving cooperatively in the initial flattening and can be obtained from (\ref{eq:04}). In this manner, we can obtain a correlation length for the breakup of particle-laden droplets by coupling the effects of the initial flattening and later stretching. The correlation length represents the scale of concentration fluctuations caused by the particle heterogeneity, and controls the stability of the suspension after the flattening and stretching process. The capillary length ${{l}_{c}}$ is
\begin{equation}\label{eq:07}
 {{l}_{c}}=\sqrt{\frac{\sigma }{{{\rho }_{f}}a}},
\end{equation}
where $a$ is the stretching acceleration of the droplet caused by the airflow and can be estimated as
\begin{equation}\label{eq:08}
  a\sim \frac{\frac{1}{2}{{C}_{d}}{{\rho }_{g}}u_{g}^{2} (\frac{\pi }{4}d_{w}^{2})}{\frac{\pi }{6}d_{0}^{3}{{\rho }_{d}}}=\frac{3{{C}_{d}}}{4}\frac{{{\rho }_{g}}u_{g}^{2}d_{w}^{2}}{{{\rho }_{d}}d_{0}^{3}},
\end{equation}
where $C_d$ is the drag coefficient and can be considered as a constant $C_d = 1.2$ for simplicity \citep{White2003FluidMechanics}. $d_w$ is the width of the droplet after the initial flattening and can be estimated as ${{d}_{w}}\sim {{d}_{0}}\We_{c,0}^{1/2}$ \citep{Xu2023Viscosity}. By substituting (\ref{eq:04}), (\ref{eq:07}) and (\ref{eq:08}) into (\ref{eq:06}), we can obtain a dimensionless correlation length for the breakup of particle-laden droplets
\begin{equation}\label{eq:09}
 \frac{\xi }{{{d}_{0}}}\sim {{\left( 4\phi _{m}^{2}/3{{C}_{d}}\We_{c,0}^{2} \right)}^{1/3}}{{\left( {{\phi }_{m}}-\phi  \right)}^{-5/6}}{{\left( \frac{{{d}_{p}}}{{{d}_{0}}} \right)}^{1/3}}.
\end{equation}

From (\ref{eq:09}), we can know that the correlation length is mainly controlled by the volume fraction $\phi $ and the particle size ${{d}_{p}}$. Based on this correlation length, we will further analyse the fragment size distribution in two typical breakup modes, including the bag breakup and the low-order multimode breakup. As for the shear-stripping breakup, the fragments consist mainly of individual particles and mists of the interstitial liquid. Since the particle size is known, and the mists are too small to measure, the fragment size in the shear-stripping breakup mode will not be analysed here.

\subsubsection{Fragments in bag breakup mode}\label{sec:3.5.1}

\begin{figure}
  \centerline{\includegraphics[width=0.95\columnwidth]{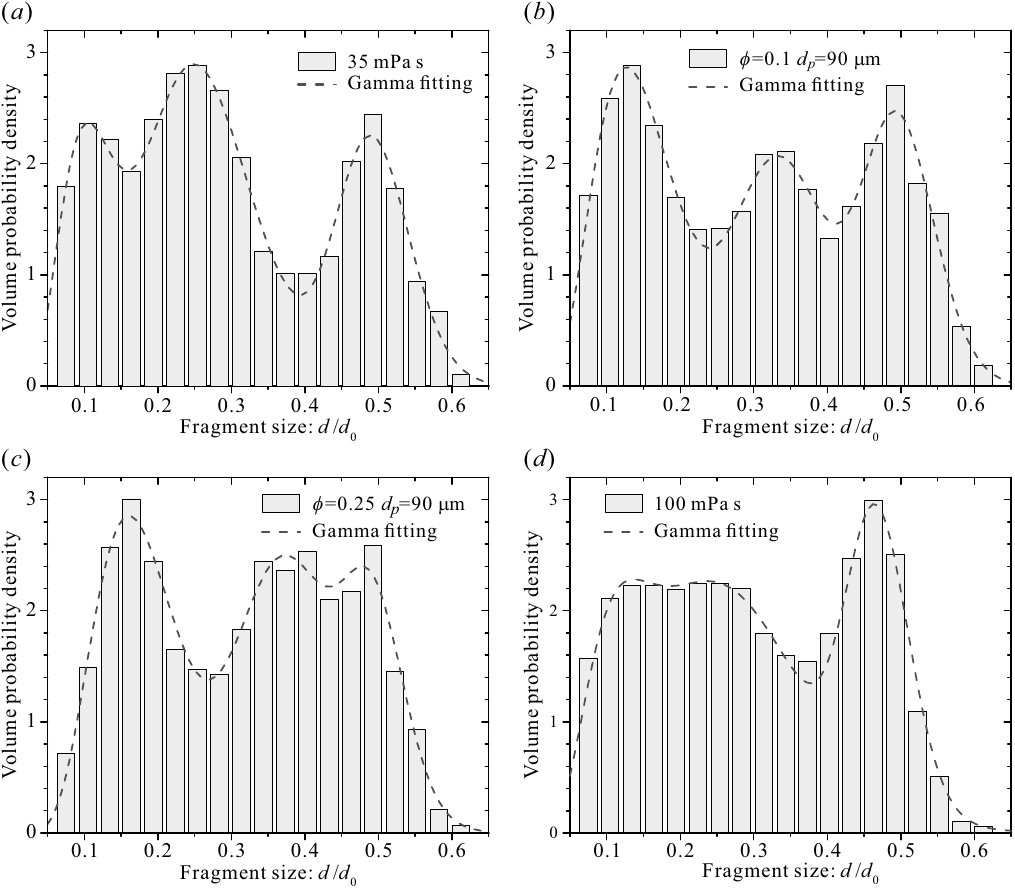}}
  \caption{Volume probability density distribution of the fragments in the bag breakup. The corresponding cases are the stars along the dashed line with $\We_{c,0} = 12.2$ in figure \ref{fig:fig09}: \ff{a} silicone oil with $\eta = 35$ mPa s, $\We_g = 15.7$, $\Oh = 0.11$; \ff{b} suspension with $\phi = 0.10$, $\We_g = 16.6$, $\Oh = 0.17$; \ff{c} suspension with $\phi = 0.25$, $\We_g = 20.3$, $\Oh = 0.32$; \ff{d} silicone oil with $\eta = 100$ mPa s, $\We_g = 20.3$, $\Oh = 0.32$. $d_0 = 3.35$ mm. The dashed lines are the compound gamma distributions with the best weight fitting. Each plot shows the statistical results of about 60 repeated experiments under the same conditions.}
\label{fig:fig13}
\end{figure}

To quantify the fragment size, we use the probability density of fragment size weighted by volume, which represents the ratio of the volume of fragments with a certain diameter to the total volume. As shown in figure \ref{fig:fig13}, these distributions of volume probability density in the bag breakup mode contain three distinct peaks. Combined with the morphological evolution of bag breakup, we can know that the three peaks, from small to large, correspond to the fragmentation of different parts of the droplet, including the bag, ring, and nodes. Compared with silicone oil droplets, the fragment size distributions of the particle-laden droplets show distinct differences. The positions of the two peaks for small droplets shift to the right, while the position of the peak for large droplets remains almost unchanged. In addition, the heights of different peaks also change, corresponding to the variation of the volume fractions of different parts.

To further analyse the effect of particles on the fragment size distribution, we need to distinguish the effects of particles on different parts of the droplet (i.e., the bag, ring, and nodes). For each part, the fragments essentially come from the rupture of threads or ligaments, so the fragment size distribution corresponding to each part of the droplet can be fitted with a gamma distribution \citep{Villermaux2020FragmentationCohesion, Xu2022ShearBreakup}. The parameters of the gamma distribution corresponding to different parts depend on the average fragment size and the corrugation of the ligaments from which the fragments originate \citep{Villermaux2020FragmentationCohesion, Villermaux2004Gamma}. Hence, the parameters of the gamma distribution are different for different parts. The different gamma distributions for different parts can be combined to form a compound gamma distribution. The overall size distribution is fitted by this compound gamma distribution, and different peaks of the overall size distribution correspond to the different sub-gamma distributions \citep{Jackiw2022SizeDistribution, Someshwar2023SizeDistribution}. In this regard, we use a compound gamma distribution to fit the overall size distribution
\begin{equation}\label{eq:10}
 g\left( x \right) = \sum\limits_{i = 1}^3 {{w_i}{g_{i,v}}\left( x \right)},
\end{equation}
where $x=d/{{d}_{0}}$ is the dimensionless fragment size and $d_0$ is the initial diameter of the droplet, ${{w}_{i}}$ is the weight of each part and $w_1 + w_2 + w_3=1$, ${g_{i,v}}\left( x \right) = {\beta _{i}}^{{\alpha _i}}{x^{{\alpha _i} - 1}}{e^{ - {\beta _i}x}}/\Gamma \left( {{\alpha _i}} \right)$ is the volume probability density function for each part, $\Gamma $ is the gamma function, ${{\alpha }_{i}}$ and ${{\beta }_{i}}$ are the shape parameter and the rate parameter of ${g_{i,v}}$, respectively. Since the compound gamma distribution contains many parameters, a direct fitting using the gamma distribution may obtain different results and thus make quantitative comparisons difficult. Here, for the fragment size distribution of the bag breakup, the interaction between the peaks is small due to the large difference in the size of the fragments produced by different parts. So we can determine ${\alpha _i}$ and ${\beta _i}$ of each sub-gamma distribution based on the shape and position of each peak, and then obtain the weight ${w_i}$ of each peak by fitting. In this way, the compound gamma distributions are in good agreement with the measured results, as shown by the dashed lines in figure \ref{fig:fig13}. Based on the parameters of each volume probability density function, we can obtain the shape parameter and the rate parameters of each number probability density function as ${\alpha _i} - 3$ and ${\beta _i}$ \citep{Jackiw2022SizeDistribution}, and further estimate the average fragment size of the corresponding part as
\begin{equation}\label{eq:11}
  {d_i} = \frac{{{\alpha _i} - 3}}{{{\beta _i}}},{\rm{ }}i = 1,2,3,
\end{equation}
where ${{d}_{1}}$, ${{d}_{2}}$, and ${{d}_{3}}$ are the average fragment sizes corresponding to the bag (${{d}_{b}}$), ring (${{d}_{r}}$), and node (${{d}_{n}}$), respectively. Moreover, the weight ${{w}_{i}}$ of each sub-gamma distribution is the volume fraction of that part of the fragments.

\begin{figure}
  \centerline{\includegraphics[width=1\columnwidth]{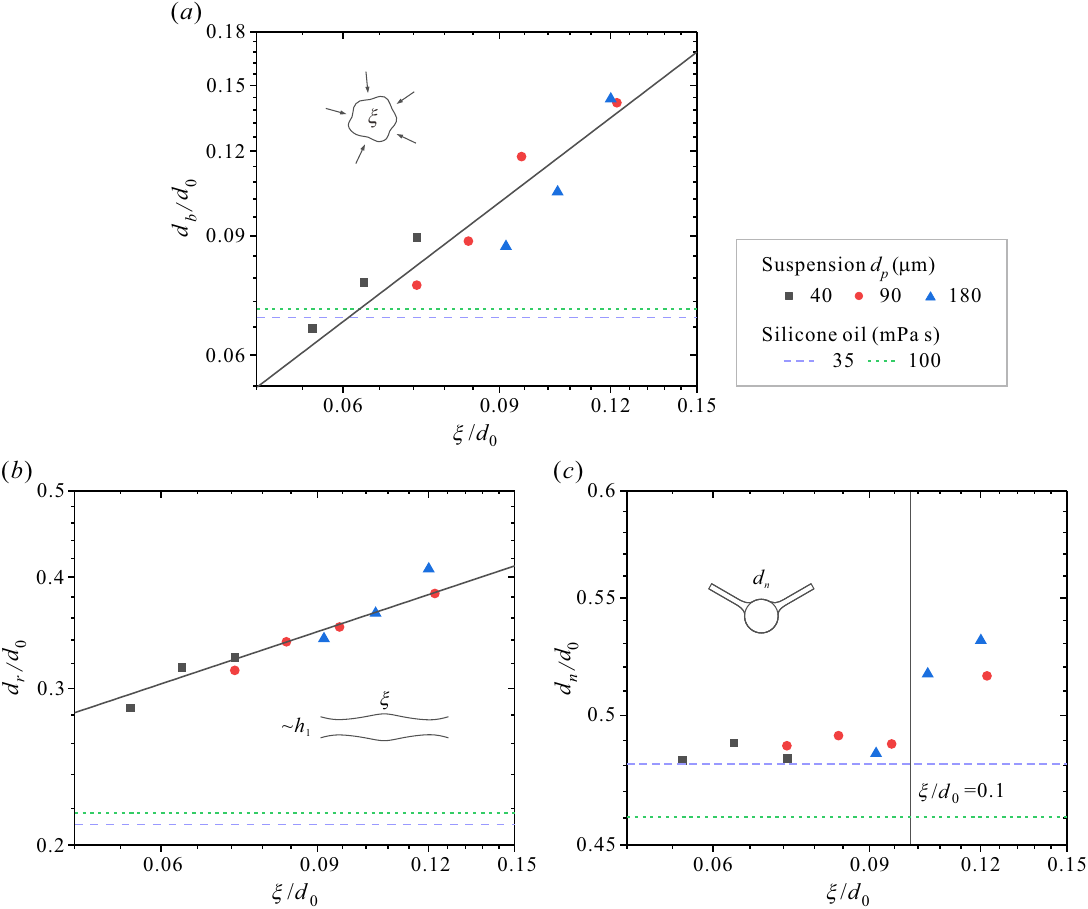}}
  \caption{Average fragment sizes of \ff{a} the bag, \ff{b} the ring, and \ff{c} the node under different particle volume fractions and particle sizes. The scatters are the fragment size of the particle-laden droplets obtained from (\ref{eq:11}). The dashed and dotted lines correspond to the fragment sizes of silicone oil droplets with $\eta$ = 35 and 100 mPa s, respectively. $\xi /{{d}_{0}}$ is calculated according to (\ref{eq:09}), $d_0 = 3.35$ mm. The corresponding cases are the stars along the dashed line with $\We_{c,0} = 12.2$ in figure \ref{fig:fig09}. Each data point shows the statistical results of about 60 repeated experiments under the same condition. The insets illustrate the fragment mechanisms for the corresponding parts.}
\label{fig:fig14}
\end{figure}

The average fragment size corresponding to different parts of the droplet obtained by (\ref{eq:11}) is shown in figure \ref{fig:fig14}. The fragment size of particle-laden droplets is generally larger than that of silicone oil droplets, and the scaling relations between the correlation length and the fragment size are different for different parts. For the fragmentation of the bag, the liquid film retracts towards particle clusters to form the fragments, as illustrated in the inset of figure \ref{fig:fig14}\ff{a}. The sizes of the particle clusters can be characterised by the correlation length. This is because the particle clusters essentially correspond to the concentration fluctuation (i.e., the particle agglomeration or separation), and the coherence length reflects the range of concentration fluctuation caused by the heterogeneous effect of particles. Hence, the fragment size of the bag ($d_b$) can be estimated as
\begin{equation}\label{eq:12}
  \frac{{{d}_{b}}}{{{d}_{0}}}\sim \frac{\xi }{{{d}_{0}}}.
\end{equation}
The relation is in good agreement with the data in figure \ref{fig:fig14}\ff{a}, where the scaling in (\ref{eq:12}) is shown as the solid line with a prefactor of $C_1 = 1.12$.

For the fragmentation of the ring, each fragment corresponds to a segment of the liquid ring, as illustrated in the inset of figure \ref{fig:fig14}\ff{b}. The initial diameter of the liquid ring, which is approximately equal to the thickness of the droplet ($h$) after the initial ﬂattening, is larger than the correlation length ($\xi $). But as the liquid ring is stretched and thinned, the diameter of the liquid ring ($h_1$) is smaller than the correlation length. At this condition, the concentration fluctuation caused by the particle heterogeneity can lead to the destabilization of the ring and then produce fragments. Therefore, the correlation length reflecting the range of concentration fluctuation can characterise the wavelength of the destabilization, which is the length of the segment of the liquid ring where a fragment is produced. In addition, the diameter of the liquid ring ($h_1$) is mainly controlled by the rim expansion ratio, which is affected by the Weber number \citep{Jackiw2021InternalFlow}. Hence, $h_1$ can be considered as a constant due to the same inviscid Weber number in our experiments. Therefore, the fragment size of the ring can be estimated based on the mass conservation as
\begin{equation}\label{eq:13}
  \frac{{{d_r}}}{{{d_0}}} \sim {\left( {\frac{\xi }{{{d_0}}}} \right)^{1/3}}.
\end{equation}
The relation is in good agreement with the data in figure \ref{fig:fig14}\ff{b}, where the scaling in (\ref{eq:13}) is shown as the solid line with a prefactor of $C_2 = 0.78$.

Finally, for the fragmentation of the nodes, the fragment size is almost constant unless the correlation length $\xi $ is large ($\xi /{{d}_{0}}>0.1$), as shown in figure \ref{fig:fig14}\ff{c}. This is because the effect of particle heterogeneity occurs when the characteristic scale is smaller than the correlation length. For the nodes with a large scale, the size of the node fragments will be affected only when the correlation length is large, that is, a high volume fraction or a large particle size. For particle-laden droplets with $\xi /{{d}_{0}}<0.1$, we find that the fragment size of the node obeys
\begin{equation}\label{eq:14}
  \frac{{{d}_{n}}}{{{d}_{0}}}={{C}_{3}}\frac{h}{{{d}_{0}}}.
\end{equation}
where the thickness of the droplet after the initial flattening $h$ is a constant for the cases in this study, and $C_3$ is a constant equal to 1.55. In figure \ref{fig:fig14}\ff{c}, the dashed line for silicone oil droplets with $\eta$ = 35 mPa s corresponds to $C_3 = 1.53$ ($d_n / d_0 = 0.481$), and the dotted line for silicone oil droplets with $\eta = 100$ mPa s corresponds to $C_3 = 1.47$ ($d_n / d_0 = 0.460$). This law is consistent with the end-pinching mechanism proposed by \cite{Schulkes1996LiquidFilaments} for free liquid jets (with $C_3 = 1.55$), that is, the droplets are shed via end-pinching caused by the retraction of ligament tips, as illustrated in the inset of figure \ref{fig:fig14}\ff{c}. A similar law was also found in the breakup of Worthington jets \citep{Gordillo2010JFM} and unsteady sheets \citep{Wang2018SheetFragmentation}, which have $C_3 =$ 1.5--1.6. Here, our results show that the fragmentation of the node also follows this scaling law for the end-pinching mechanism.

\begin{figure}
  \centerline{\includegraphics[width=0.95\columnwidth]{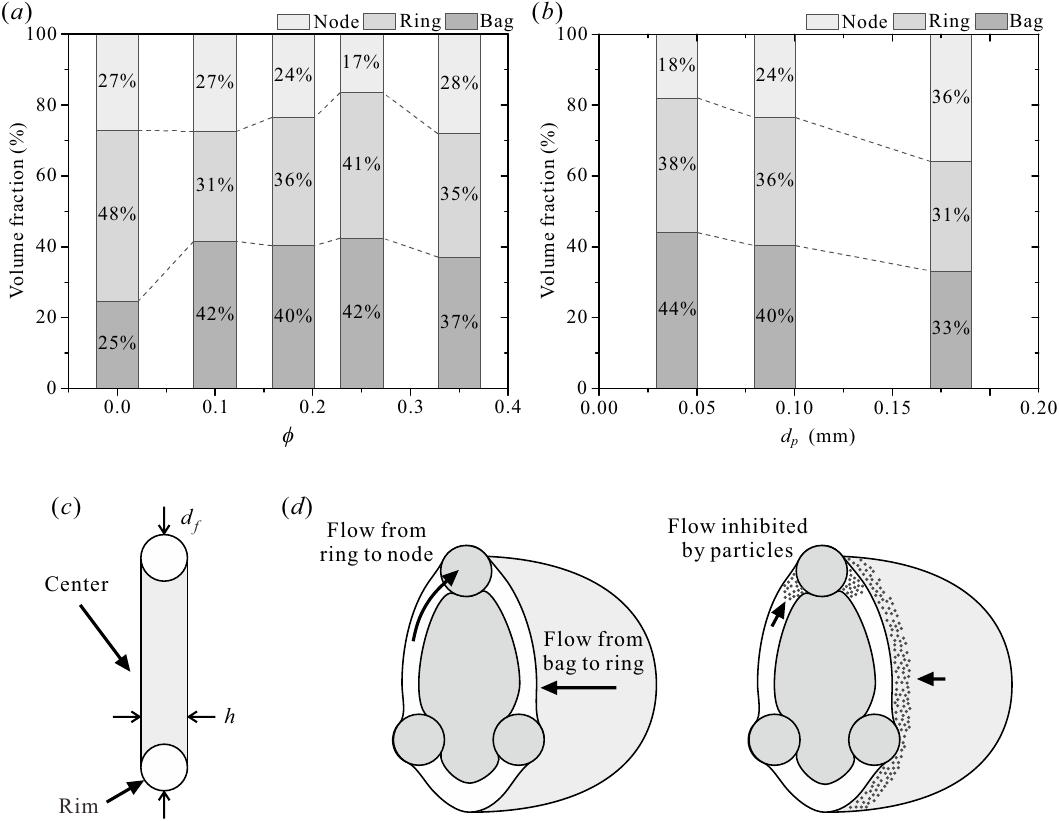}}
  \caption{Variation of the volume fraction of each part with \ff{a} the particle volume fraction and \ff{b} the particle size. The corresponding cases are the stars along the dashed line with $\We_{c,0}$ = 12.2 in figure \ref{fig:fig09}. Each data point shows the statistical results of about 60 repeated experiments under the same conditions. \ff{c} Illustration of the droplet morphology after the initial flattening. \ff{d} Illustration of the liquid flow in the bag development process and the effect of particles on flow inhibition.}
\label{fig:fig15}
\end{figure}

In addition to the average fragment size, the volume fraction of each part is also important for understanding the fragment size distribution. The volume fraction of each part obtained by the weight ${{w}_{i}}$ of each sub-gamma distribution is shown in figures \ref{fig:fig15}\ff{a-b}. To analyse the reason of the volume fraction of each part, we need to consider the mass distribution in the droplet after the initial flattening. This is because the evolution of the droplet shape and the subsequent breakup process determined the volume fraction of each part. After the initial flattening, the droplet is deformed into a disk with a thickness of $h\approx 1.05$ mm and a width of ${{d}_{f}}\approx 5.6$ mm, as shown in figure \ref{fig:fig15}\ff{c}. Correspondingly, the volume fractions at the rim and middle of the droplet are roughly 56\% and 44\%, respectively (based on volume calculation from figure \ref{fig:fig15}\ff{c}). Then the middle of the droplet develops into the bag, and the rim develops into the ring and nodes. During this process, liquid flows from the bag to the ring and from the ring to the nodes \citep{Jackiw2022SizeDistribution}, as illustrated in figure \ref{fig:fig15}\ff{d}. These processes ultimately determine the volume fraction of each part. For the silicone oil droplet ($\phi $ = 0 in figure \ref{fig:fig15}\ff{a}), the flow from the bag to the ring reduces the volume fraction of the middle bag from the initial 44\% to 25\% after the breakup. The volume fraction of the nodes calculated from the fragment size (\ref{eq:14}) and number (usually about 3) is 28\%, which is in agreement with the experimental data (27\%) in figure \ref{fig:fig15}\ff{a}. In addition, \cite{Jackiw2022SizeDistribution} predicted the fragment size distribution of water droplets, and found the volume fraction of the bag, ring and nodes were 12\%, 45\% and 38\%, respectively. Compared with their prediction, the volume fraction of the bag in our results is larger, while the volume fraction of the node is smaller. Besides the effect of a slight difference in $\We_{c,0}$, this difference may be due to the higher viscosity of the droplet in our experiments, which inhibits the liquid flow between the parts. These comparisons also indicate that the volume fractions obtained from the compound gamma distribution are reasonable.

Compared with the silicone oil droplet, there are several differences in the volume fractions of different parts of the particle-laden droplet. First, the volume fraction of the bag of the particle-laden droplet is much larger, i.e., about 40\%. This is because, during the bag development process, the particles tend to be retained at the junction of the bag film and the ring. This results in a higher local particle concentration and correspondingly a higher viscosity, which inhibits the liquid flow from the bag to the ring, as illustrated in figure \ref{fig:fig15}\ff{d}. In addition, the particle clusters on the liquid film can trap some liquid. Therefore, the bag volume fraction of the particle-laden droplet is maintained at about 40\%, i.e., the initial volume fraction in the middle of the droplet. Second, as the particle volume fraction increases, the volume fraction of the nodes decreases first and then increases, while the volume fraction of the ring increases first and then decreases, as shown in figure \ref{fig:fig15}\ff{a}. This is because the higher particle volume fraction inhibits the liquid flow from the ring to the nodes. However, when the particle volume fraction is very large ($\phi $ = 0.35 in figure \ref{fig:fig15}\ff{a}), the presence of the particles makes the fragment size of the node increase significantly (as shown in figure \ref{fig:fig14}\ff{c}), and the corresponding volume fraction of the node also increases. Third, as the particle size increases, the volume fraction of the nodes increases gradually, and the volume fractions of the bag and ring decrease, as shown in figure \ref{fig:fig15}\ff{b}. This is because the larger particle tends to reside on the larger-scale nodes rather than the smaller-scale bag and ring.

\begin{figure}
  \centerline{\includegraphics[width=0.7\columnwidth]{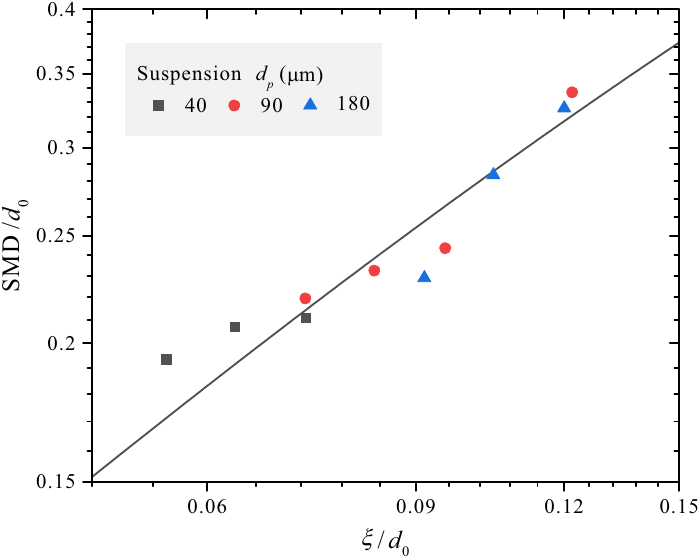}}
  \caption{Sauter mean diameter (SMD) of fragments at different particle volume fractions and sizes in the bag breakup mode. The solid line is (\ref{eq:16}) with a prefactor of 1.4. The corresponding cases are the stars along the dashed line with $\We_{c,0}$ = 12.2 in figure \ref{fig:fig09}. $\xi /{{d}_{0}}$ is calculated according to (\ref{eq:09}), $d_0 = 3.35$ mm. Each data point shows the statistical results of about 60 repeated experiments under the same conditions.}
\label{fig:fig16}
\end{figure}

In addition to considering the effects of particles on different parts, the fragment size distribution can also be described by some representative diameter to quantify the overall atomisation effect. A representative diameter is the Sauter mean diameter (SMD), which is the diameter of a droplet whose ratio of volume to surface area is the same as that of the entire fragments \citep{Lefebvre2017Atomization}. The SMD can be calculated as
\begin{equation}\label{eq:15}
  \text{SMD}=\frac{\sum{{{n}_{i}}d_{i}^{3}}}{\sum{{{n}_{i}}d_{i}^{2}}}=\frac{1}{\sum{{{{w}_{i}}}/{{{d}_{i}}}}},
\end{equation}
where $n_i$ is the number of fragments corresponding to a size of $d_i$ and equal to ${{{w}_{i}}d_{0}^{3}}/{d_{i}^{3}}$. Combining the fragment sizes (obtained by (\ref{eq:12}), (\ref{eq:13}) and (\ref{eq:14})) and the corresponding volume fractions of different parts, we can estimate the SMD as
\begin{equation}\label{eq:16}
 \frac{{{\text{SMD}}}}{{{d_0}}} \sim \frac{{1/{d_0}}}{{{w_1}/{d_b} + {w_2}/{d_r} + {w_3}/{d_n}}} = {\left[ {\frac{{{w_1}}}{{{C_1}}}{{\left( {\frac{\xi }{{{d_0}}}} \right)}^{ - 1}} + \frac{{{w_2}}}{{{C_2}}}{{\left( {\frac{\xi }{{{d_0}}}} \right)}^{ - 1/3}} + \frac{{{w_3}}}{{{C_3}}}{{\left( {\frac{h}{{{d_0}}}} \right)}^{ - 1}}} \right]^{ - 1}}.
\end{equation}
By substituting the prefactors $C_1 = 1.12$, $C_2 = 0.78$, $C_3 = 1.55$ (obtained from (\ref{eq:12}), (\ref{eq:13}) and (\ref{eq:14})) and the weights $w_1 = 0.405$, $w_2 = 0.345$, $w_3 = 0.25$ (obtained by averaging all the cases considered in this study) into (\ref{eq:16}), we can see that the relationship is in good agreement with the experimental data, as shown in figure \ref{fig:fig16}, where (\ref{eq:16}) is shown as the solid line with a prefactor of 1.4.

\subsubsection{Fragments in multimode breakup}\label{sec:3.5.2}
\begin{figure}
  \centerline{\includegraphics[width=0.95\columnwidth]{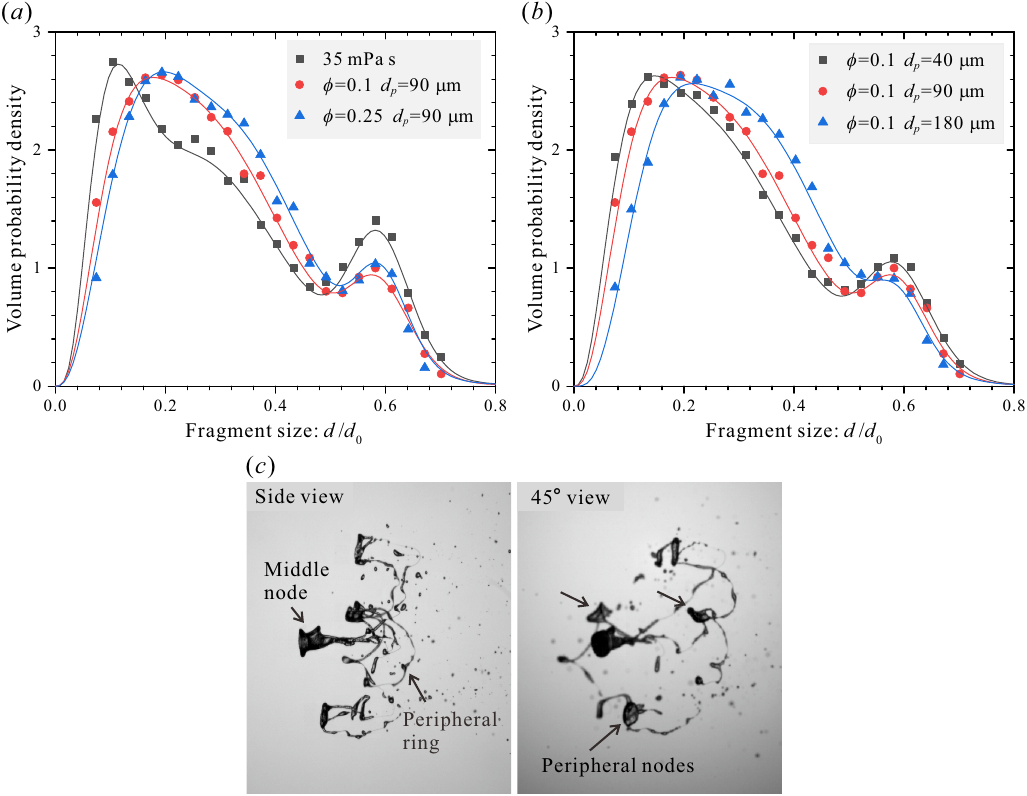}}
  \caption{Variation of the volume probability density distribution in the low-order multimode breakup with \ff{a} the particle volume fraction and \ff{b} the particle size. The corresponding cases are the stars along the dashed line with $\We_{c,0} = 36$ in figure \ref{fig:fig09}, i.e., fluids used are silicone oil with $\eta$ = 35 mPa s, $\We_g = 51.6$, $\Oh = 0.11$, and suspensions with $\phi = 0.10$, $\We_g = 56.3$, $\Oh = 0.17$ and $\phi = 0.25$, $\We_g = 71.7$, $\Oh = 0.32$. $d_0 = 3.35$ mm. The solid lines are the compound gamma distributions with the best weight fitting. Each plot shows the statistical results of about 90 repeated experiments under the same conditions. \ff{c}~Droplet morphology at 14.5 ms in low-order multimode breakup mode, corresponding to the whole breakup process shown in figure \ref{fig:fig05}.}
\label{fig:fig17}
\end{figure}

The volume probability density distribution of the fragments in the low-order multimode breakup is shown in figure \ref{fig:fig17}. The fragments are also formed by the fragmentation of three parts, including the middle node, the peripheral nodes and the peripheral ring, as illustrated in figure \ref{fig:fig17}\ff{c}. The fragmentation of the middle node forms a distinct peak with the largest fragment size. But the fragment sizes formed by the fragmentation of the peripheral nodes and ring are close, and the corresponding peaks are indistinguishable. Therefore, we still use the compound of three gamma distributions to fit the data, but do not distinguish the effects of particles on different parts. We only use the fitting curves to identify the changes in the overall size distributions. As shown in figures \ref{fig:fig17}\ff{a-b}, the fragment size distribution shifts to the right as the particle volume fraction and particle size increase, that is, the overall fragment size of the particle-laden droplets becomes larger.

\begin{figure}
  \centerline{\includegraphics[width=0.7\columnwidth]{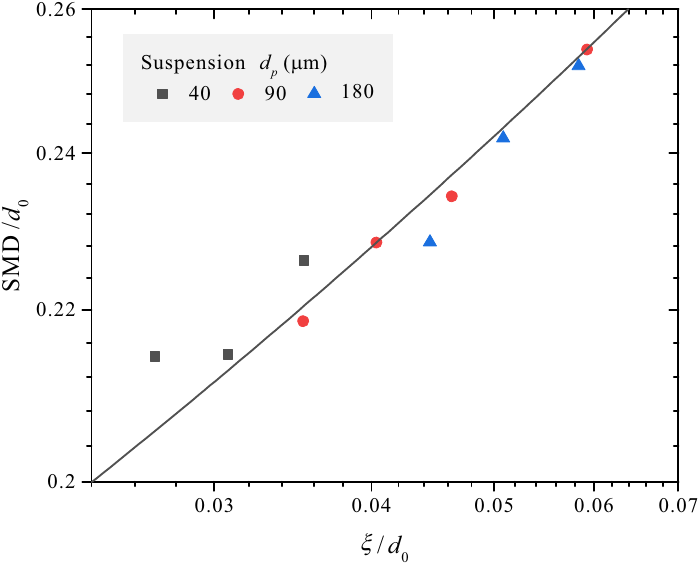}}
  \caption{Sauter mean diameter (SMD) of fragments at different particle volume fractions and particle sizes in the low-order multimode breakup. The corresponding cases are the stars along the dashed line with $\We_{c,0}$ = 36 in figure \ref{fig:fig09}. $\xi /{{d}_{0}}$ is calculated according to (\ref{eq:09}), $d_0 = 3.35$ mm. The solid line is (\ref{eq:20}) with $C_s = 0.764$. Each point shows the statistical results of about 90 repeated experiments under the same condition.}
\label{fig:fig18}
\end{figure}

The Sauter mean diameters of the fragment size distributions are shown in figure \ref{fig:fig18}. For the multimode breakup, the droplet is pierced by RT instability waves to form fragments with a typical size ($d_s$). The breakup is extremely unstable, and the heterogeneity of the particle-laden fluid will intensify the breakup randomness, as described in \S \ref{sec:3.4}. For this condition, we can refer to the full-wave approach to predict the Sauter mean diameter. The full-wave approach assumes that there is a spectrum of wavelengths which could be excited to appreciable amplitudes and then produce fragments of corresponding size \citep{Mayer1961Atomization, Yang2018FullWaves, Yang2020FullWaves}. Following this idea, in the condition of the multimode breakup, the fragments originate from the piercing caused by the RT instability wave, so the sizes of the fragments ($d_s$) can be estimated as
\begin{equation}\label{eq:17}
  {{d}_{s}}={{C}_{s}}{{\lambda }_{RT}},
\end{equation}
where $C_s$ is a constant and ${{\lambda }_{RT}}$ is the RT instability wavelength. In addition, the number probability of fragments with size $d_s$ is required to calculate SMD. However, due to the close size of the fragments originating from different parts, the exact size distribution of fragments is difficult to be quantified by the compound gamma distributions. Since we mainly focus on the effect of particles, we, for simplicity, neglect the variation of the volume probability density with the fragment size $d_s$, that is, the volume of the initial droplet is divided uniformly into $N$ parts and each of which is broken into fragments with size $d_s$. Therefore, the number probability ($n$) of fragments with size $d_s$ is estimated as
\begin{equation}\label{eq:18}
  n \sim \frac{{d_0^3}}{{Nd_s^3}}.
\end{equation}
By integrating over the spectrum of wavelengths and combining (\ref{eq:17}) and (\ref{eq:18}), the SMD can be written as
\begin{equation}\label{eq:19}
  \text{SMD}=\frac{\int_{{{\lambda }_{\min }}}^{{{\lambda }_{\max }}}{nd_{s}^{3}d\lambda }}{\int_{{{\lambda }_{\min }}}^{{{\lambda }_{\max }}}{nd_{s}^{2}d\lambda }}={{C}_{s}}\frac{{{\lambda }_{\max }}-{{\lambda }_{\min }}}{\ln {{\lambda }_{\max }}-\ln {{\lambda }_{\min }}},
\end{equation}
where ${{\lambda }_{\max }}$ and ${{\lambda }_{\min }}$ are the upper and lower limits of the spectrum of wavelengths. For droplet breakup, the wavelength must be smaller than the droplet diameter, i.e., ${{\lambda }_{\max }}={{d}_{0}}$. The correlation length of the particle-laden droplet is the typical length over which a fluctuation will be damped, so short-wavelength instabilities will be suppressed by the particles, i.e., ${{\lambda }_{\min }}=\xi $. Therefore, the SMD in (\ref{eq:19}) can be written in dimensionless form as
\begin{equation}\label{eq:20}
  \frac{\text{SMD}}{{{d}_{0}}}={{C}_{s}}\frac{1-\xi /{{d}_{0}}}{-\ln \left( \xi /{{d}_{0}} \right)}.
\end{equation}
Through the full-wave approach with a lower limit $\xi $, the relation in (\ref{eq:20}) reflects the effect of the particle in intensifying the breakup randomness and suppressing the short-wavelength instabilities, and is in good agreement with the data in figure \ref{fig:fig18}, where (\ref{eq:20}) is the solid line with $C_s = 0.764$.

\section{Conclusions}\label{sec:4}
In this study, we investigate the breakup of particle-laden droplets in airflow experimentally. By combining high-speed images from the side view and the 45$^\circ$ view, the morphologies of particle-laden droplets and homogeneous fluid droplets are compared in detail. In the bag breakup mode, the particles agglomerate on the liquid film and accelerate the breakup of the liquid film. In the multimode breakup, the particles induce localised rapid piercing. In the shear-stripping breakup, the breakup process is more discrete, and the particles and interstitial liquid are almost completely separated. A breakup regime map is produced, and transitions between different breakup modes are obtained based on a unified model (\ref{eq:02}). The effect of particles on the transition of breakup modes is manifested by a higher effective viscosity compared to the interstitial fluid.

Furthermore, to quantitatively evaluate the effects of particles, we eliminate the effect of the higher effective viscosity for particle-laden droplets by comparing the cases with the same $\We_{c,0}$. For the stretching and fragmentation of the liquid film, the particles accelerate the breakup of the film due to the less liquid content in the film and the local stress caused by the particles. For the piercing in the droplet middle, the particles induce local piercing due to the change in effective viscosity when the droplet thickness is reduced close to the particle size, which further leads to more abundant and random piercing phenomena of the particle-laden droplet.

Finally, considering the coupling effects of the initial flattening and later stretching in the droplet breakup process, we propose a correlation length ($\xi$) that controls the stability of the suspension through a concentration fluctuation. Based on the correlation length, the fragment size distributions in the bag breakup and low-order multimode breakup are analysed theoretically. For the bag breakup, the fragment sizes corresponding to different parts of the droplet obey different scaling relations with the correlation length, and the volume fractions of different parts also change with the particle diameter and the volume fraction. By combining the fragment sizes and the corresponding volume fractions of different parts, the SMD of fragments of the particle-laden droplet in the bag breakup mode is obtained. For the multimode breakup, the overall fragment size of the particle-laden droplet is larger, and its SMD can be predicted by the full-wave approach with the correlation length as the lower limit.

\section*{Supplementary data}\label{SupMat}Supplementary material and movies are available online.
\section*{Acknowledgements}This work was supported by the National Natural Science Foundation of China (Grant Nos.\ 51676137, 52176083, 51920105010, and 51921004).
\section*{Declaration of interests}
The authors report no conflict of interest.
\section*{Author ORCID}Zhizhao Che, https://orcid.org/0000-0002-0682-0603

\appendix
\section{Rheological measurements of the suspensions}\label{sec:AppA}
\begin{figure}
  \centerline{\includegraphics[width=0.62\columnwidth]{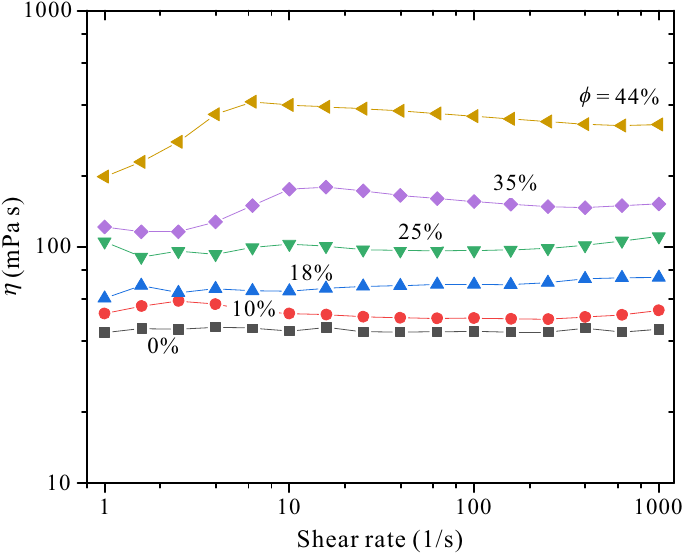}}
  \caption{Viscosity of the suspensions with different particle volume fractions ($\phi$).}
\label{fig:fig19}
\end{figure}

The rheology of the suspensions was measured using a rotational rheometer (Anton Paar MCR 501) with a rough parallel plate. The plates had a diameter of 25 mm and the gap between the plates was 1 mm. Before the measurements, the suspension was pre-sheared at 0.01 rps for three minutes. During the measurements, each data point was obtained by maintaining the corresponding shear rate for 10 s. The test temperature was controlled at 22 $^\circ$C. The results are presented in figure \ref{fig:fig19}. As the particle volume fraction increases, the suspension gradually exhibits non-Newtonian properties at low shear rates. But at a high shear rate, the suspensions with different particle volume fractions maintain an almost constant viscosity. For droplet breakup in the airflow, the droplet experiences a high shear rate. It can be estimated by the characteristic shear rate $\dot{\gamma }={{u}_{g}}\sqrt{{{\rho }_{g}}/{{\rho }_{f}}}/{{d}_{0}}$ \citep{Che2023ShearThinningBreakup}, which is in the range of 90--700 s$^{-1}$ in our experiments. Therefore, we used the averaged viscosity at shear rates of 100--1000 s$^{-1}$ as the effective viscosity of the suspensions.

\bibliographystyle{jfm}
\bibliography{ParticleDropletBreakup}
\end{document}